\documentclass[preprint]{revtex4}
\usepackage{graphicx}
\usepackage{dcolumn}
\usepackage{amsmath}
\usepackage[latin1]{inputenc}
\usepackage{graphicx, psfrag}
\usepackage{amssymb}
\usepackage[colorlinks=true, citecolor=blue, urlcolor = blue, linkcolor= blue, bookmarks=true]{hyperref}
\usepackage{float}
\usepackage{amsmath}
\usepackage{amsfonts}
\usepackage{dcolumn}
\usepackage{hyperref}
\usepackage{subfigure}
\usepackage{pgfplots}
\usepackage{epstopdf}

\makeatletter
\def\btt#1{\texttt{\@backslashchar#1}}
\DeclareRobustCommand\bblash{\btt{\@backslashchar}} \makeatother

\makeatletter
\def\btt#1{\texttt{\@backslashchar#1}}
\DeclareRobustCommand\bblash{\btt{\@backslashchar}} \makeatother

\begin{document}	
\title[]{Gravitational lensing for stationary axisymmetric black holes in Eddington-inspired Born-Infeld gravity}

\author{Md Sabir Ali} 
\email{sabir.ali@iitrpr.ac.in} 
\affiliation{Indian Institute of Technology Ropar, Punjab-140001, India}
\author{Shagun Kauhsal}
\email{2018phz0006@iitrpr.ac.in} 
\affiliation{Indian Institute of Technology Ropar, Punjab-140001, India}

\begin{abstract}
\noindent	
The recent years witnessed a surge of interest of the lensing of the black holes arising from general as well as other modified theories of gravity due to the experimental data available from the EHT results. The EHT may open a new door indicating the possible existence of the rotating black hole solutions in modified theories of gravity in the strong field regime. With this motivation,  we investigate in the present paper the equatorial lensing $(\theta=\pi/2)$ by a recently obtained exact rotating black holes solution in EiBI theory in both the strong and weak field limits. Such black holes are the modification of Kerr-Newman black holes in general relativity, characterized by their mass ($M$), the charge ($Q$), and the rotation parameter ($a$). and an additional term $\epsilon$ accounting for the correction to the Kerr-Newman solutions. We show numerically the variations of the impact parameter $u_m$, the light deflection coefficients $p$ and $q$, the total azimuthal bending angle $\alpha_D$ and find a close dependence of these quantities on the charge parameter $r_q$, the correction term $\epsilon$ and the spin $a$. We also calculate the angular position $\theta_\infty$, and the angular separation $s$, and the magnification of the relativistic images.  In addition, we also discuss the weak lensing of the black holes in EiBI theory using the Gauss-Bonnet theorem. We calculate the weak lensing parameter  and find its variation with different values of the parameters $r_q$ and $\epsilon$.
\end{abstract}
\maketitle

\newpage

\tableofcontents
\section{Introduction}
Einstein's general relativity admits only a few exact physically acceptable solutions \cite{ExactBook}. Among these limited solutions, we have the axially symmetric stationary solutions, such as Kerr and Kerr-Newman which are, respectively, the vacuum and electrovacuum solutions of general relativity \cite{UT1, UT2, UT3, Kerr:1963ud, Newman:1965my}. These solutions are well established in the context of no-hair theorems for the asymptotically flat axisymmetric spacetimes. However, in the presence of hair, no-hair theorems may find another solutions which are characterized by some other parameters. An extension of this possibility has been done in the case of scalar and proca hairs in the asymptotically de sitter spacetimes \cite{Bhattacharya:2011dq}. It is also shown in another investigation that the no-hair theorem ruled out the possible existence of the real massive vector field in $f(R)$ theories as long as the potential due to scalar field is positive definite in Einstein's frame \cite{Bhattacharya:2016lup}. To check the Kerr black holes hypothesis to know the exact nature of the astrophysical black holes is a proven fact that has been tested many times using X-ray spectroscopy of the accreting matter around the black holes \cite{Jiang:2014loa, Cardenas-Avendano:2016zml}, the strong gravitational lensing, and the recently obtained images of black holes silhouettes of M87 supermassive black holes using the Event Horizon Telescope (EHT) \cite{Akiyama:2019cqa, Akiyama:2019brx, Akiyama:2019sww, Akiyama:2019bqs, Akiyama:2019fyp, Akiyama:2019eap}. Apart from these, gravitational wave astronomy confirms that the gravitational waves emerging from black hole mergers using Kerr solution are strongly matches with the waveform signals as detected by LIGO Scientific-Virgo Collaborations \cite{Abbott:2016blz, Abbott:2017oio}. But, these results may not find their validity in the strong gravity regime, and the alternative theories of gravity may find a new resource to test the strong gravity field with the more advanced future technologies \cite{Abbott:2018lct}. The awakening of the gravitational wave astronomy by LIGO/Virgo and the imaging of the shadow of M87 black holes by EHT may open a new door to the physics world, particularly, in the strong gravity regime such as black hole physics. Since the astrophysical black holes are mostly rotating, thereby obtaining the exact rotating solutions in different gravity theories is a pressing topic to test the strong gravity regime through gravitational waves, and imaging of black holes mostly residing presumably at the heart of the every galaxies.    

Having obtained a black hole solution, it is worthy of investigating one of the most prominent astrophysical events that occurred when light rays pass through such compact objects. The light rays or any massive particles while encountered with the black holes, their direction of propagation drastically changes. Light rays instead of following the straight path, would follow the curved path what that black holes have created in their surroundings. Such phenomenon is called the gravitational lensing. After its first-ever imaging of the silhouettes of the M87 supermassive black holes by the EHT, the study of the lensing phenomena in the strong gravity region has been of utmost importance. The shadow imaging in the sky relies on the gravitational lensing of strong field as encompassed by light rays, thereby bearing the fingerprints of the geometry of the strong field gravity. After its first inception by Darwin \cite{Darwin}, the studies on the lens equation and lensing phenomena of the astrophysical compact objects were triggered. Later Frittelli, Kling and Newman \cite{Frittelli:1999yf}, and then Virbhadra \cite{Virbhadra:1999nm}, analyzed the lens equation without referring to the black holes background. After that, the lens equation by Virbhadra and Ellis for the Schwarzschild black holes was constructed. In the subsequent years, Bozza \textit{et al.} \cite{Bozza:2001xd, Bozza:2002zj} following the Virbhadra-Ellis lens equation invented a mathematical formulation of the gravitational lensing of a generic black hole in a spherically symmetric spacetime. Motivated by the formulation and with the advance of time, people made tremendous development in the investigation of the strong gravitational lensing of the various static spherically symmetric as well as axially symmetric stationary black holes spacetimes \cite{Will, Bhadra:2003zs, Whisker:2004gq, Eiroa:2004gh, Eiroa:2005vd, Keeton:2005jd, Sarkar:2006ry, Keeton:2006sa, Iyer:2006cn, Zhang:2007nk, Chen:2009eu, Reyes:2010tr,Bozza:2007gt, Wei:2011nj, Eiroa:2010wm, Sadeghi}. The light bending angle is studied to rule out the possible inhomogeneity in the ideal the dark energy fluid, which is contained within a cosmic structure \cite{Ali:2017ofu}. The bending angle and also the perihilon precission of light are also investigated for the spacetimes in Hordeski gravity theories in the realm of astrophysical scenarios \cite{Bhattacharya:2016naa}. The gravitational lensing by black holes have been investigated using both analytical and numerical techniques \cite{Virbhadra:2007kw, Man, Chen:2009eu, Sarkar:2006ry, Javed:2019qyg, Shaikh:2019itn, Stefanov:2010xz, Cardoso:2008bp, Hod:2009td, Kumar:2019ku, Kumar:2020ju, Kumar:2020hgm, Kumar:2020sag, Islam:2020xmy, Islam:2020sag, Ohanian:1987pc, Einstein,Virbhadra:2002ju,Gavazzi:2008,Bozza:2002af,Virbhadra:1999nm}. 
With the advent of modern technologies, the EHT group has been able to image the black holes silhouettes using very long baseline interferometry (VLBI) techniques \cite{Akiyama:2019cqa, Akiyama:2019brx, Akiyama:2019sww, Akiyama:2019bqs, Akiyama:2019fyp, Akiyama:2019eap}. They observed the very first image of the shadow of the M87 supermassive black holes by modelling the Kerr spacetime. These investigations from EHT placed a strong piece of evidence at first sight that there could be no other spacetimes apart from the Kerr metric \cite{Akiyama:2019cqa}. However, recently measured values of the rotation parameter show uncertainties to what could have expected using the Kerr metric as a source \cite{Bambi1}. Therefore, there should be imposed some minimal constraint conditions on the angular measurement of the Kerr black holes \cite{Psaltis:2020lvx}. Therefore, the non-Kerr black holes showing the significant deviations in measurement procedures, cannot be a possible candidate from the phenomenological point of view \cite{Psaltis:2020lvx}.

There have been a lot of research interests in the investigations of the strong gravitational lensing for a non-Kerr family of black holes. The physical observables for the lensing effect in the strong domain of gravity have been investigated for various rotating non-Kerr spacetimes, e.g., the hairy Kerr black holes \cite{Islam:2021dyk}, the nonsingular Kerr-Sen black holes \cite{Kumar:2020sag}, the rotating black holes in $4D$ EGB gravity 
\cite{Ghosh:2020spb}, etc. They studied rigorously various observables, such as the light deflection angle, the angular distance, and the angular separation and angular magnifications, and also the time delay effects to investigate the astrophysical consequences in the context of the black holes M87 and SgrA$^*$ \cite{Guerrero:2020azx}. Motivated by these ideas, in this paper we aim to discuss these physical observables for strong lensing for the rotating solution in the context of EiBI gravity theory. This is possible to test the strong field gravitational effects using such a non-Kerr family of black holes for a variety of observations.  

The paper is organized as follows. In the Sec.~\ref{Sec2}, we briefly review the rotating black holes in Eddington-inspired Born-Infeld gravity. The usual formalism to derive the strong lensing observable we refer to Sec.~\ref{Sec3}. The numerical techniques and plots of light deflection angle, the angular distance and the angular separations are obtained in  Sec.~\ref{Sec4}. The derivations of the weak field light bending angle using Gauss-Bonnet theorem is the subject of Sec.~\ref{weak_lensing}. We conclude the paper in the Sec.~\ref{Sec6}.\\

\section{EiBI gravity and rotating solutions}\label{Sec2}
The rotating solutions in the Eddington-inspired Born-Infeld gravity are obtained when one employs the correspondence between modified models as a contraction of a metric tensor with the Ricci scalars formulated in the light of general relativity and the Ricci-based gravity theories. For the basic investigations and the properties of the rotating black holes in EiBI gravity theories we refer our reader to \cite{Guerrero:2020azx} and the references therein. This is an exact solution that is obtained when there is a nonminimally coupling to nonlinear electrodynamics of Born-Infeld gravity. The spacetime is a rotating  charged black holes in EiBI gravity theories when written in the usual Boyer-Lindquist coordinates ($t,x,\theta, \phi$) reads as \cite{Guerrero:2020azx}
	\begin{eqnarray} \label{eq:axialline}
	ds^2&=&-\left(1-f +\epsilon \rho^q \frac{(\Delta+ a^2 \sin ^2\theta)}{\Sigma}\right) dt^2
	- 2 a \left(f-\epsilon \rho^q\frac{ (\Delta+x^2+a^2)}{\Sigma}\right) \sin ^2\theta dt d\phi   +\frac{(1+\epsilon \rho^q)\Sigma}{\Delta} dx^2\nonumber\\
	&+&(1-\epsilon \rho^q) \Sigma d\theta^2
	+\Big[ \left(x^2+a^2+f a^2\sin ^2\theta\right)
	-\epsilon \rho^q \frac{(x^2+a^2)^2+a^2\Delta \sin^2\theta}{\Sigma} \Big]   \sin ^2\theta d\phi^2,
	\end{eqnarray}
where the correction term $\epsilon$ encodes the deviation of charged rotating black holes in EiBI gravity to that of the Kerr-Newman metric of general relativity. We have also noted that
\begin{eqnarray}
\label{eq:defsBL}
f&=& \frac{r_S x-r_q^2/2}{\Sigma}=\frac{x^2+ a^2-\Delta}{\Sigma} \nonumber \\
\Sigma&=& x^2+ a^2\cos^2\theta \nonumber \\
\Delta&=& x^2-r_S x+a^2+r_q^2/2,
\end{eqnarray}
and 
\begin{equation}\label{eq:densityKN}
\rho^q= \frac{r_q^2}{2\Sigma^2} \ , 
\end{equation}
which can be viewed as the energy density of a charged rotating black holes whose electromagnetic field is described by
\begin{equation}
A_{\mu}=(A_t,0,0,A_{\phi})=\frac{Qx}{\Sigma}(1,0,0,-a \sin^2 \theta) \ ,
\end{equation}
from which we can immediately get the required components of the field strength tensor $F_{\mu\nu}=\nabla_\mu A_{\nu}-\nabla_\nu A_{\mu}$. The quantity $a$ is the spin angular momentum of the black holes and $r_q$ is the charge parameter. The rotating black holes (\ref{eq:axialline}) in EiBI gravity encompasses the Kerr-Newman black hole when $\epsilon=0$ and Kerr black holes in the case of $r_q=0$. To calculate the various observables in the study of the strong lensing, we are in a position to introduce the following dimensionless quantities   
\begin{eqnarray}
x \to \frac{x}{r_S},\; a \to \frac{a}{r_S}, \;t \to \frac{t}{r_S},\;r_q \to \frac{r_q}{r_S},
\end{eqnarray}
with this the metric (\ref{eq:axialline}) is recast as
\begin{eqnarray} \label{eq:axialline1}
	ds^2&&=-\left(1-\tilde{f} +\epsilon \tilde{\rho^q} \frac{(\tilde{\Delta}+ a^2 \sin ^2\theta)}{\tilde{\Sigma}}\right) dt^2
	- 2 a \left(\tilde{f}-\epsilon \tilde{\rho^q}\frac{ (\tilde{\Delta}+x^2+a^2)}{\tilde{\Sigma}}\right) \sin ^2\theta dt d\phi+\frac{(1+\epsilon \tilde{\rho^q})\tilde{\Sigma}}{\tilde{\Delta}} dx^2\nonumber\\
	&&+(1-\epsilon \tilde{\rho^q}) \tilde{\Sigma d}\theta^2+\Big[ \left(x^2+a^2+\tilde{f} x^2\sin ^2\theta\right)
	-\epsilon \tilde{\rho^q} \frac{(x^2+a^2)^2+a^2\tilde{\Delta} \sin^2\theta}{\tilde{\Sigma}} \Big] \sin ^2\theta\;d\phi^2
	\end{eqnarray}
We have also noted that
\begin{eqnarray}
\label{eq:defsBL}
\tilde{f}&=& \frac{x-r_q^2/2}{\tilde{\Sigma}}=\frac{x^2+a^2-\tilde{\Delta}}{\tilde{\Sigma}} \nonumber \\
\tilde{\Sigma}&=& x^2+a^2\cos^2\theta \nonumber\\
\tilde{\Delta}&=& x^2-x+a^2+r_q^2/2.
\end{eqnarray}
and 
\begin{equation}\label{eq:densityKN}
\tilde{\rho}^q= \frac{r_q^2}{2\tilde{\Sigma^2}} \ , 
\end{equation}
The rotating black holes in EiBI gravity is a stationary axially symmetric spacetimes which is invariant under the simultaneous transformation $t\to -t$ and $\phi\to -\phi+2\pi$. Therefore, the metric (\ref{eq:axialline1}) admits two Killing vectors, 
$\eta^{\mu}_{(t)}=\delta^{\mu}_t $ and $\eta^{\mu}_{(\phi)}=\delta^{\mu}_{\phi}$ which are linearly independent. The vectors $\eta^{\mu}_{(t)}$ and $\eta^{\mu}_{(\phi)}$ are associated, respectively, with the translational and rotational isometries \cite{Chandrasekhar:1992}. The event horizon is a well defined boundary that is a null hypersurface and it comprises of the outward null geodesics which are not capable to hit the null infinity in the future. The event horizon is a solution of $g^{xx}=\tilde{\Delta}=0$, which leads to have the form
\begin{equation}
\label{horizoneq}
x_\pm=\frac{1\pm\sqrt{1-4(a^2+r_q^2/2)}}{2}.
\end{equation}
which has the same expression as of the Kerr-Newman black holes. Therefore the event horizon of the rotating black holes in EiBI black holes theory has the similar structure as of the Kerr-Newman black holes. However, the static limit surface are not same as $g_{tt}^{\text{KN}}\neq g_{tt}^{\text{EiBI}}$. 
\begin{figure}
	\begin{centering}
		\begin{tabular}{c c}
		    \includegraphics[scale=0.45]{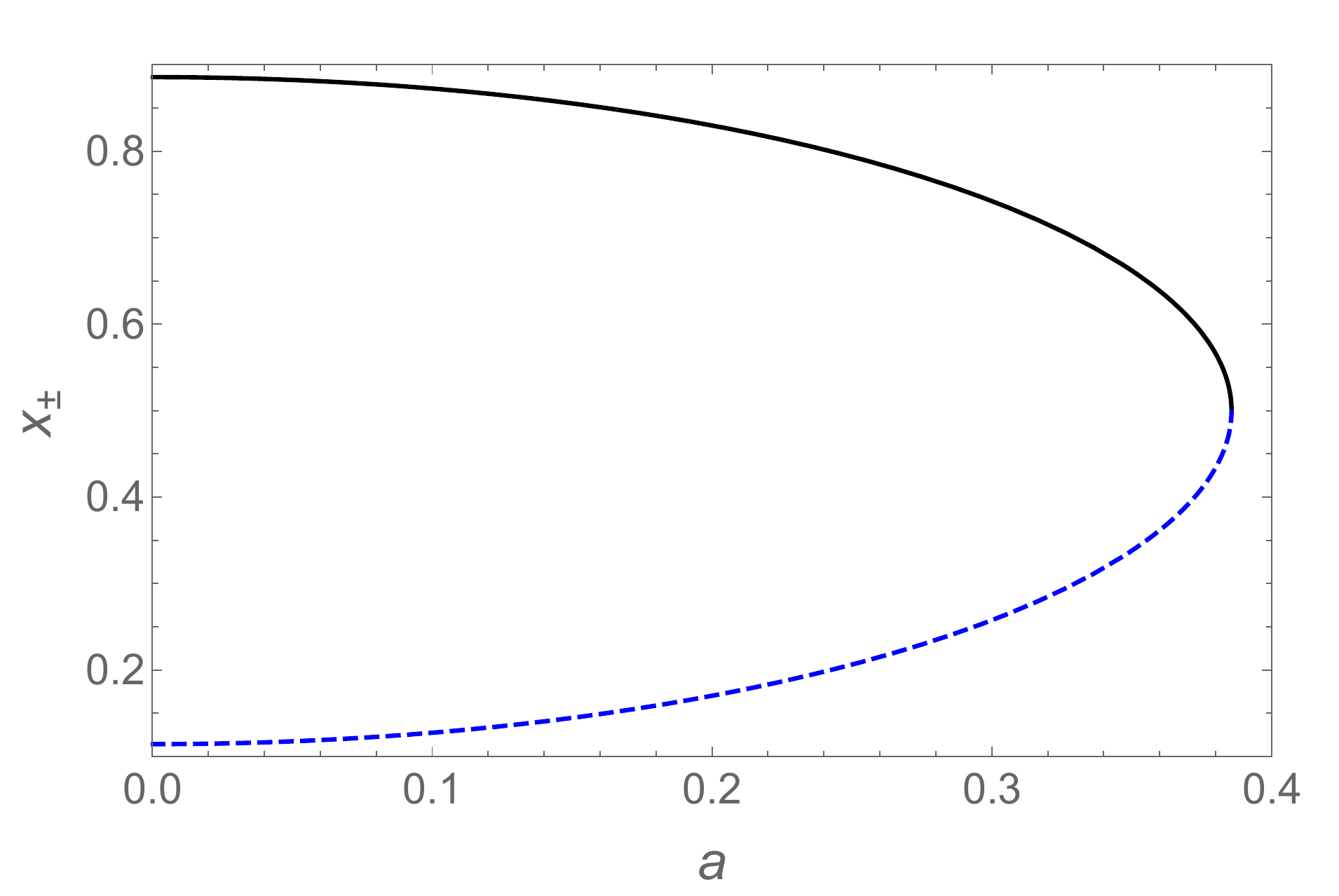}\hspace*{0.2cm}&
		    \includegraphics[scale=0.45]{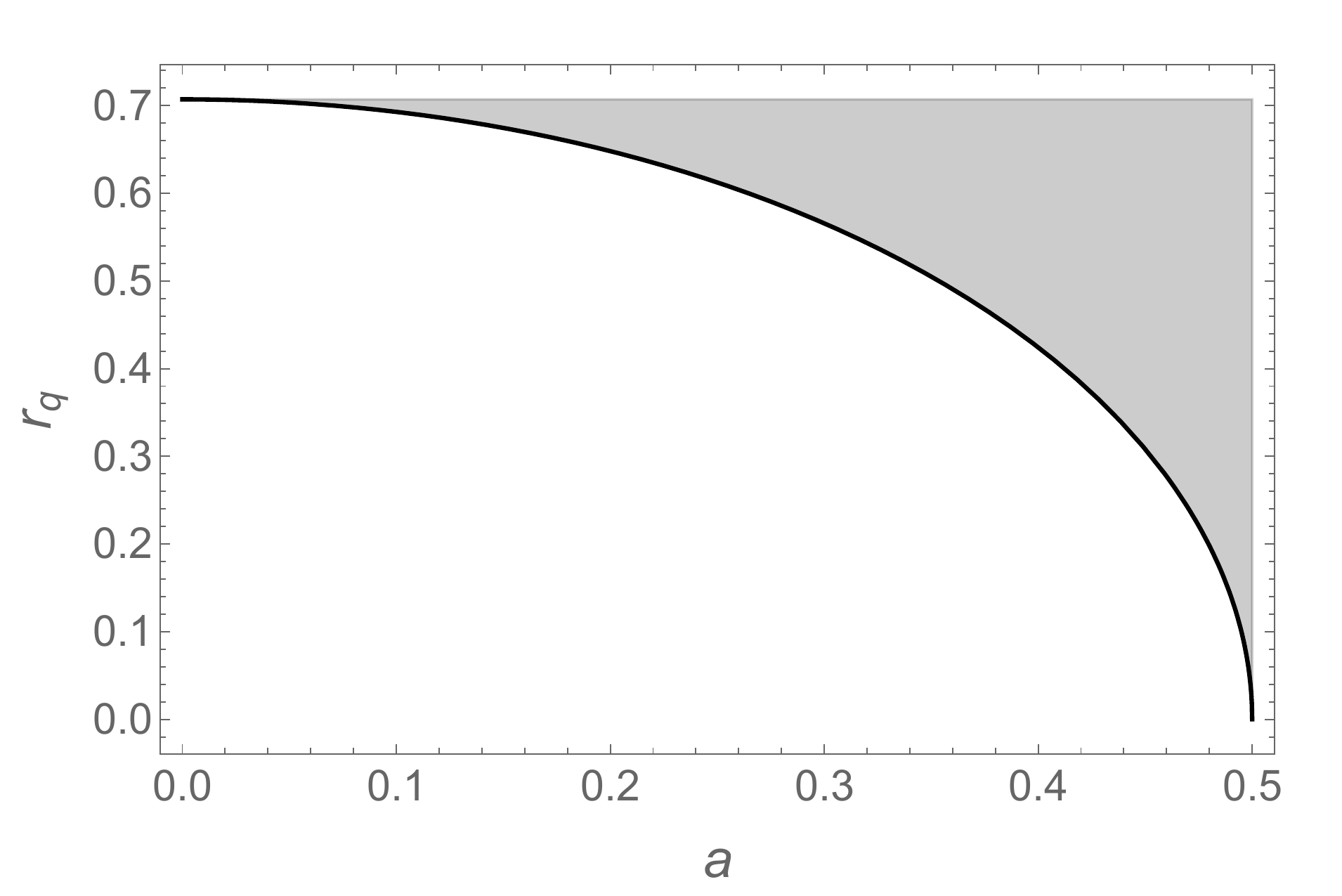}\\
			\end{tabular}
	\end{centering}
    \caption{The plot of horizon radius vs the rotation parameter for $r_q=0.45$ (left figure) and the charge parameter vs the rotation parameter (right figure).}
    \label{fig:horizon}
\end{figure}
\\
The black hole exists only in the limit when $a^2\leq 1/4-r_q^2/2$. The maximum value of the rotation parameter turns out to be $a=0.5$, for $r_q=0$. For any non-zero value of the parameter $r_q$, the rotation parameter has value less than $0.5$. The left side of Fig~\ref{fig:horizon} shows the variation of the horizon $x_\pm$ with respect to the rotation parameter $a$. The blue dotted curve represents the Cauchy horizon, $x_-$ whereas the black solid line represents the event horizon, $x_+$. Similarly, the right figure shows the parameter space of the charge parameter $r_q$ with respect to the rotation parameter $a$. The shaded region in this plot shows the no black hole spacetime. This figure shows the restricted  theoretical values of $r_q$ and $a$.
\\

Given the opportunities available for testing the alternative theories of gravity, the physical implications of such black hole solutions and the analysis of their various features is very timely in the context of astrophysical settings. The data available from various experimental set up such as the EHT, the LIGO scientific and the Virgo collaborations one can study the potential deviations of the Kerr-Newman black holes in general relativity in the study of accretion disks, strong gravitational lensing and shadows, generation of gravitational waves in binary mergers, and so on from that of the rotating solutions that arise from the EiBI gravity.

\section{Equatorial black hole lensing}\label{Sec3}
In this section, we investigate the equatorial ($\theta=\pi/2$) light bending due to the rotating black holes in EiBI gravity. The effects of the deformation parameter $\epsilon$, the charge parameter $r_q$ and the spin $a$ on the equatorial lensing will also be investigated. The metric (\ref{eq:axialline1}), for the equatorial plane reads as
\begin{eqnarray}\label{NSR}
\mathrm{ds^2}=-A(x)dt^2+B(x) dx^2 +C(x)d\phi^2-D(x)dt\,d\phi,
\end{eqnarray}
where
\begin{eqnarray}
A(x)&=& (1-\frac{1}{x}+\frac{r_q^2}{2x^2}\left(1+\frac{\epsilon}{x^2}\left(1-\frac{1}{x}+\frac{2a^2+r_q^2/2}{2x^2}\right)\right),\;\;\;\;\;
B(x)=\frac{x^2+\frac{\epsilon r_q^2}{2x^2}}{\tilde{\Delta}},\nonumber\\
C(x)&=&x^2+a^2+x-r_q^2/2
	-\frac{\epsilon r_q^2}{2x^4}\frac{(x^2+a^2)^2+a^2\tilde{\Delta}}{x^2},\nonumber\\\;\;\;
D(x)&=&2a\left(x-r_q^2/2
	-\frac{\epsilon r_q^2}{2x^4}\frac{(x^2+a^2)+\tilde{\Delta}}{x^2}\right)
\end{eqnarray}
where $\tilde{\Delta}=x^2+a^2-x+r_q^2/2$.
We write the  Lagrangian
\begin{eqnarray}
\mathcal{L}=g_{\mu\nu}\dot{x}^\mu\dot{x}^\nu,
\end{eqnarray}
which is used to find the geodesics equation. The over dot describes the derivative with respect to the affinely parametrized variable, say $\lambda$. The metric (\ref{NSR}) admits two Killing vectors due to time translation and rotation which correspond, respectively, to the constant energy $\mathcal{E}$ and the constant angular momentum $\ell$ such that
\begin{align}
\label{tphi}
2\mathcal{E}&=\frac{\mathcal{\partial{L}}}{\partial \dot{t}}=g_{tt}\dot{t}+g_{t\phi}\dot{\phi},\\
\label{tphi1}
-2\ell&=\frac{\mathcal{\partial{L}}}{\partial \dot{\phi}}=g_{t\phi}\dot{t}+g_{\phi\phi}\dot{\phi}.
\end{align}
We consider $\mathcal{E}=1$ by suitably choosing the affine parameter and identify $\ell$ as the angular momentum of the photon with respect to the black hole axis. We have four first order differential equations in the equatorial plane using equations (\ref{tphi}) and (\ref{tphi1}), and also the null geodesics conditions $ds^2=0$ as
\begin{eqnarray}
\dot{t} &=& \frac{4C-2 \ell D}{4AC + D^2},\label{tdot}\\
\dot{\theta} &=& 0,\\
\dot{\phi} &=& \frac{2D+4 A \ell}{4AC + D^2},\label{phidot} \\
\dot{x} &=& \pm 2 \sqrt{\frac{C-D\ell-A\ell^2}{B(4AC + D^2)}}.\label{xdot} 
\end{eqnarray}
Since we are interested in the study of the photon trajectories in the isolated black hole system, we can safely ignore the effect of the other celestial objects on the trajectory of the photon and can well approximate the spacetime as Minkowaskian at a large enough distance. We assume that both the source and the observer are situated at a large distance from the black holes under study. This will satisfy our purpose for studying the lensing phenomena on the equatorial plane.

Now we need to focus on the effective potential for light rays, $V_{\text{eff}}$, which they follow in the radial direction only. The effective potential follows from the relation $\dot{x}^2+V_\text{eff}(x)=0$, and is given by
\begin{eqnarray}
V_{\text{eff}}(x)&=&-\frac{4(C-D\ell-A\ell^2)}{B(4AC + D^2)}.
\end{eqnarray}
At this point it is worthy to understand that in the asymptotic limit, the photons emanating from infinity approaches the black hole event horizon at some distance $x_0$ and leaves for infinity again. The impact parameter $u=\ell/\mathcal{E}=\ell$ ($\mathcal{E}=1$), is defined in the equatorial plane. Therefore for $V_{\text{eff}} =0 $, the expression for the angular momentum $\ell$ reads
\begin{eqnarray}\label{angmom}
\ell&=&u=\frac{-D_0+\sqrt{D_0^2+4A_0C_0}}{2A_0},\nonumber\\
&=&\frac{\epsilon\left(4 a^3 {r_q}^2+a {r_q}^4+4 a {r_q}^2 {x_0}^2-2 a {r_q}^2{x_0}\right)+\sqrt{2}{x_0}^2 \sqrt{\left(2 a^2+{r_q}^2+2 ({x_0}-1){x_0}\right) \left(4{x_0}^8-{r_q}^4 \epsilon ^2\right)}}{\epsilon  \left(4 a^2 {r_q}^2+{r_q}^4+2 {r_q}^2 {x_0}^2-2 {r_q}^2 {x_0}\right)+2 {r_q}^2{x_0}^4+4 ({x_0}-1) {x_0}^5}\nonumber\\
&+&\frac{2 a {r_q}^2{x_0}^4-4 a{x_0}^5}{\epsilon  \left(4 a^2 {r_q}^2+{r_q}^4+2 {r_q}^2 {x_0}^2-2 {r_q}^2 {x_0}\right)+2 {r_q}^2{x_0}^4+4 ({x_0}-1) {x_0}^5}
\end{eqnarray}
 Hence the expression for the impact parameter $u$ can be obtained once we get the expression for $x_0$. The ``+'' sign in front of the square root is meant for $a>0$, which indicates the prograde motion for light rays and for $a<0$ we have retrograde motion. The light deflection angle in a generic stationary axisymmetric spacetime for $x_0$ is expressed as
\begin{eqnarray}\label{bending1}
\alpha_{D}(x_0)=I(x_0)-\pi,
\end{eqnarray} 
where the total azimuthal angle $I(x_0)$ reads
\begin{eqnarray}\label{bending2}
I(x_0) = 2 \int_{x_0}^{\infty}\frac{d\phi}{dx} dx
= 2\int_{x_0}^{\infty}P_1(x,x_0)P_2(x,x_0)dx,
\end{eqnarray}
\begin{eqnarray}\label{bending3}
P_1(x,x_0)&=&\frac{\sqrt{B}\left(2A_0AL+A_0D\right)}{\sqrt{CA_0}\sqrt{4AC+D^2}},\nonumber\\
P_2(x,x_0)&=&\frac{1}{\sqrt{A_0-A\frac{C_0}{C}+\frac{L}{C}\left(AD_0-A_0D\right)}}.
\end{eqnarray}
\begin{figure}
	\begin{centering}
		\begin{tabular}{c c}
		    \includegraphics[scale=0.47]{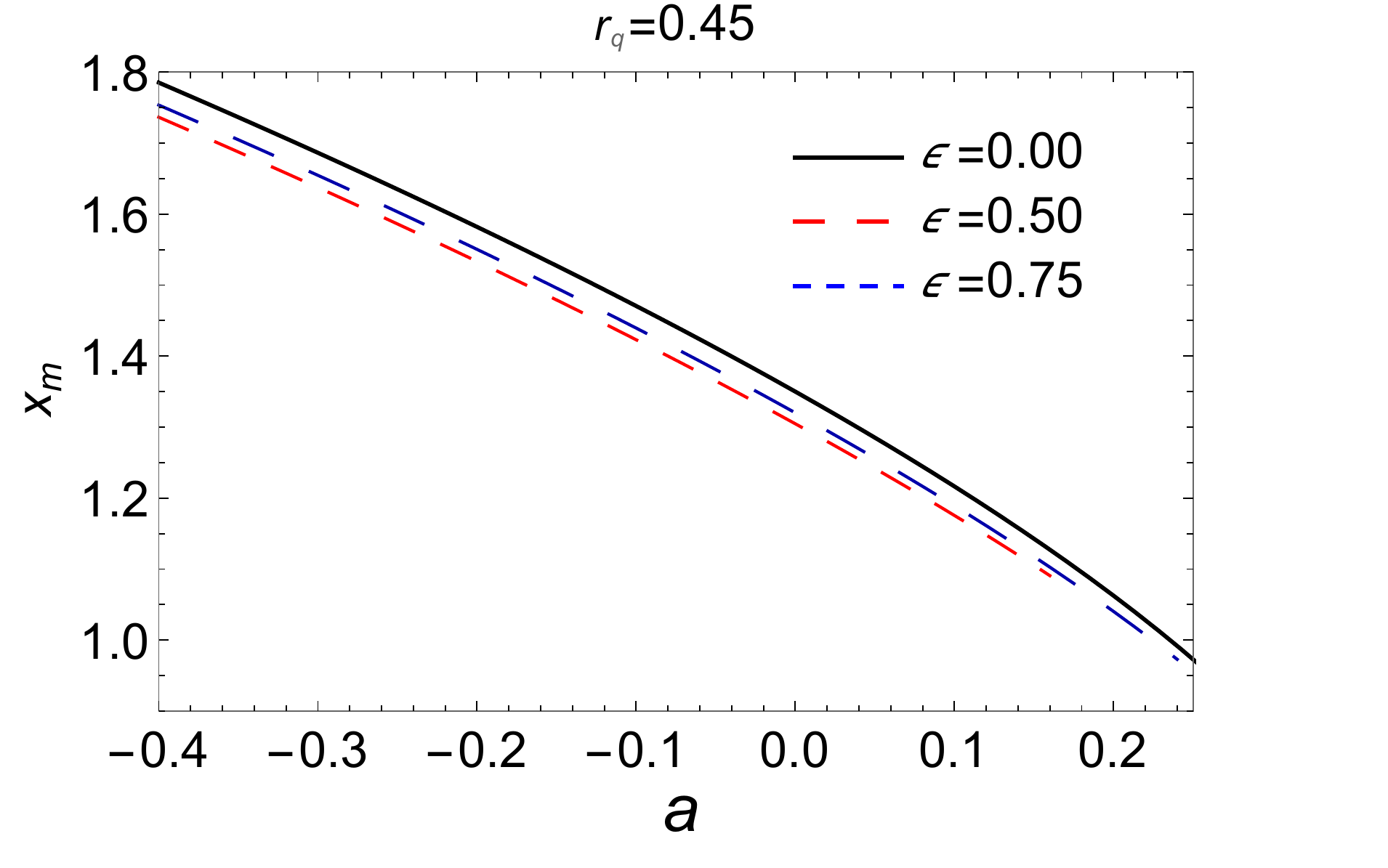}\hspace*{-0.9cm}&
		    \includegraphics[scale=0.47]{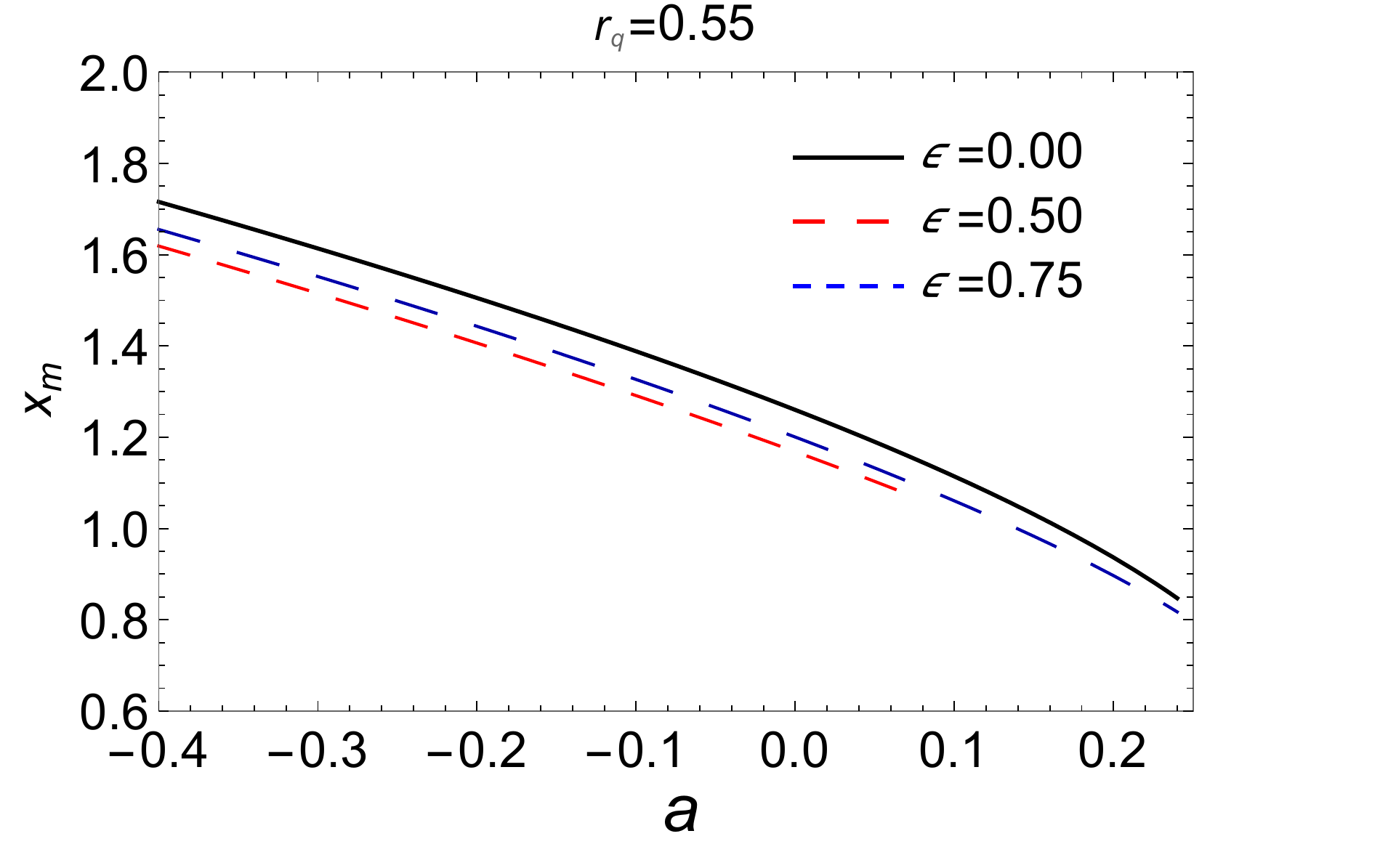}\\
		     \includegraphics[scale=0.47]{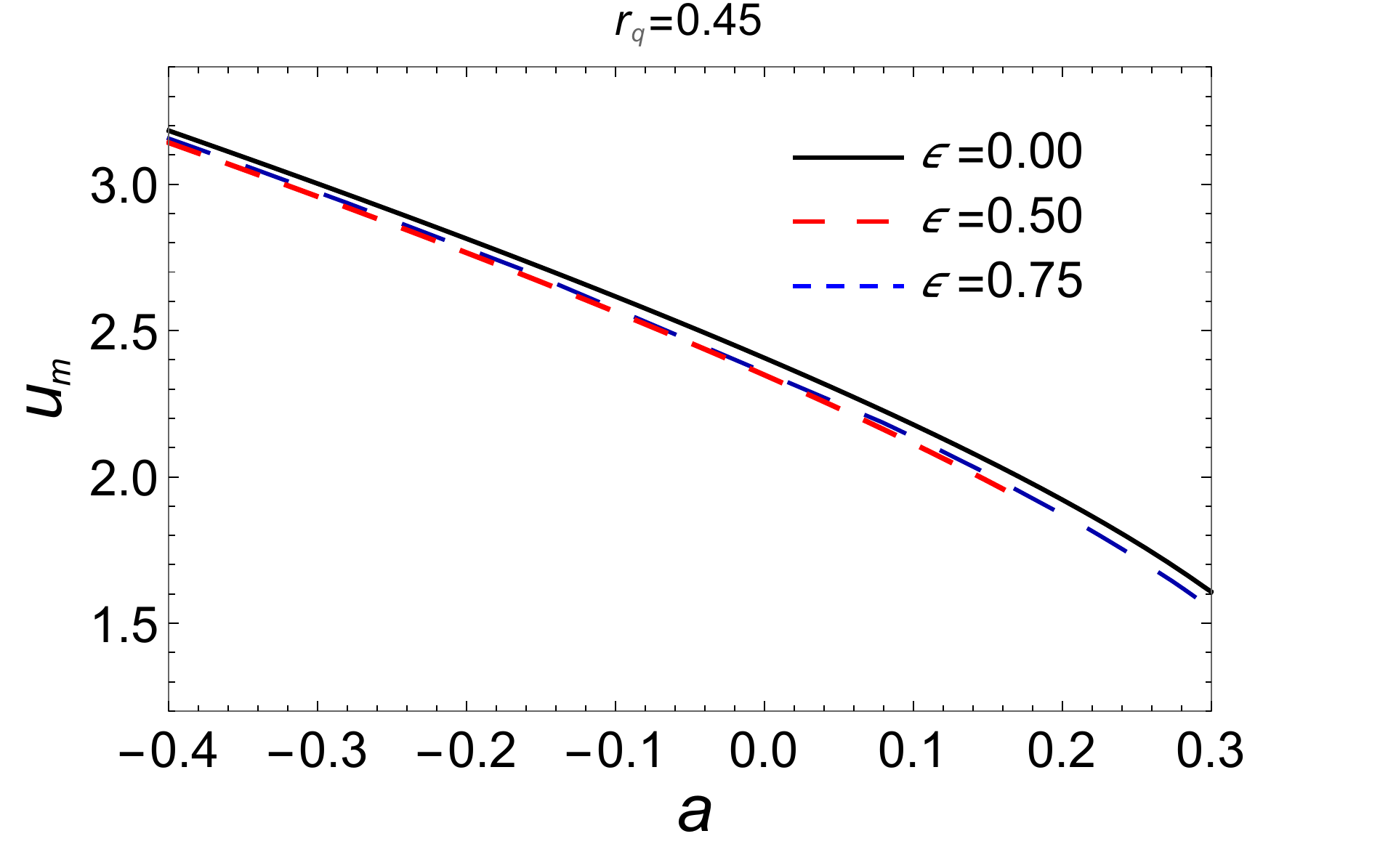}\hspace*{-0.9cm}&
		    \includegraphics[scale=0.47]{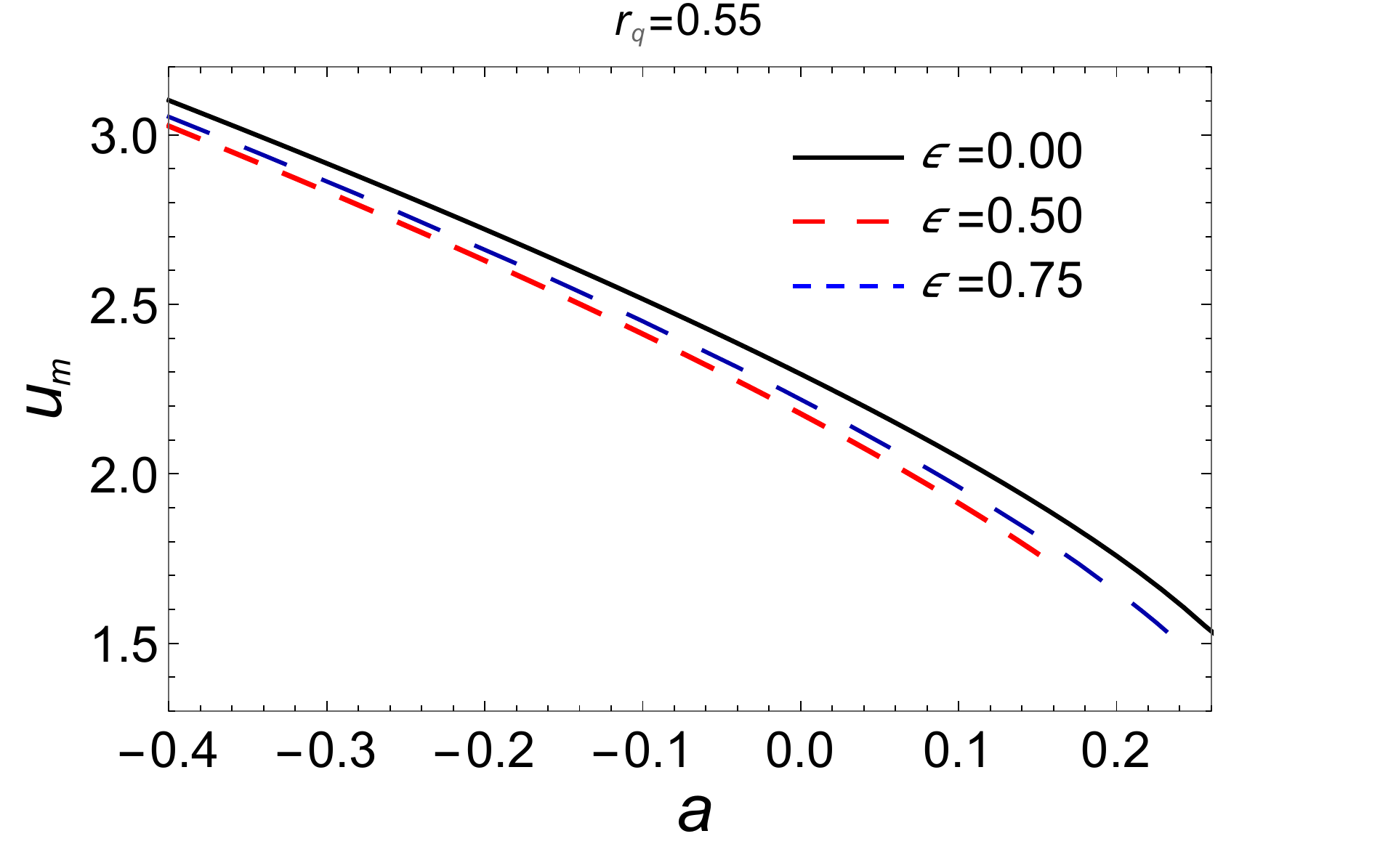}\\
		 \end{tabular}
	\end{centering}
	\caption{Plot showing the variation of the photon orbit radius $x_m$ (upper panel) and the impact parameter $u_m$ (lower panel) with respect to the rotation parameter $a$ for different values of $r_q$ and $\epsilon$.}\label{plot3}
\end{figure}  
The light rays follows a straight line along the geodesics when no black hole is present, thereby indicating $I(x_0)=\pi$. For a specific value of $x_0$ one can get $\alpha_{D}(x_0)=2\pi$, which means that the light rays would complete a whole circular loop. It goes on decreasing which eventually leads to form more than one complete loop and at a certain radius, say, $x_0 = x_m$ the total azimuthal deflection becomes infinitely large and the light rays will be impinged into the black hole. This quantity $x_m$ is called the unstable light rays' circular radius. An explicit expression for the integral (\ref{bending2}) is not obtained. Therefore, following the method as developed by Bozza \cite{Bozza:2002zj}, we calculate the behaviour of the deflection angle near the unstable photon orbit radius. In this respect, we define a new variable to separate the divergent and regular parts in $I(x_0)$, such that \cite{Bozza:2002zj}
\begin{eqnarray}
z=1 - \frac{x_0}{x}.
\end{eqnarray}
With this definition the quantity $I(x_0)$, is now expressed as
\begin{eqnarray}\label{integral}
I(x_0)=\int_{0}^{1} R(z,x_0)f(z,x_0)dz,
\end{eqnarray}
where
\begin{eqnarray}
R(z,x_0)=\frac{2x^2}{x_0}P_1(x,x_0),
\end{eqnarray}
\begin{eqnarray}
f(z,x_0)=P_2(x,x_0).
\end{eqnarray}
The function $R(z,x_0)$ in $I(x_0)$ is non-singular in nature for any value of $z$ and $x_0$, whereas the function $f(z,x_0)$ is divergent at $z=0$. To show explicitly the nature, we can Taylor expand the denominator of the function, $f(z,x_0)$ in $z$ such that
\begin{eqnarray}
f(z,x_0)\sim f_0(z,x_0)=\frac{1}{\sqrt{\alpha z+\gamma z^2+\mathcal{O}(z^3)}},
\end{eqnarray}
where we have considered the expansion upto $z^2$ only. The parameters $\alpha$ and $\gamma$ are given as
\begin{eqnarray}\label{alpha}
\alpha&=&\frac{x_0}{C_0}\left[\left(C_0^\prime A_0 -A_0^\prime C_0\right)+L\left(A_0^\prime D_0-A_0 D_0^\prime \right)\right]\\
\gamma &=& \frac{x_0}{2 C_0^2 } \left[ 2 C_0 (A_0 C'_0 - A'_0 C_0 ) + 2 x_0 C'_0 (C_0 A'_0  - A_0 C'_0) - x_0 C_0 (C_0 A''_0 - A_0  C''_0 )\right] + \nonumber\\&& L \left[\frac{x_0^2 C'_0 (A_0 D'_0 - D_0 A'_0)}{C_0{^2}} + \frac{(x_0^2/2) (D_0 A''_0 - A_0 D''_0) + x_0 (D_0 A'_0 - A_0 D'_0))}{C_0}\right].
\end{eqnarray}
Now, we find the photon orbit radius $x_{m}$, as the largest real root of the Eq.~(\ref{alpha}), such that
\begin{eqnarray}
\label{xm}
&&-16 a^4 \left({r_q}^6 \epsilon ^3-12 {r_q}^2 x_0^8 \epsilon \right)-4 a^2 \Bigg(8 x_0^{12} \left(x_0-{r_q}^2\right)-36 {r_q}^2 x_0^8 \epsilon  \left({r_q}^2+2 (x_0-1) x_0\right)+3 {r_q}^6 \nonumber\\
&&\epsilon ^3 \left({r_q}^2+2 (x_0-1) x_0\right)-2{r_q}^4 x_0^4 \epsilon ^2 \left({r_q}^2+x_0(4x_0-3)\right)\Bigg)-\Bigg(-{r_q}^6 \epsilon ^2+{r_q}^4 x_0 \epsilon  \left(8 x_0^3+\epsilon \right)\nonumber\\
&&+4 {r_q}^2 x_0^5 \left(x_0^3+4(x_0-1) \epsilon \right)-4 x_0^9\Bigg)\left(2 \sqrt{2} a-8 \sqrt{2} a^3 {r_q}^2 \epsilon  \left(6 x_0^4-{r_q}^2 \epsilon \right)\right) \sqrt{2\tilde{\Delta_0}\left(4 x_0^8-{r_q}^4 \epsilon ^2\right)}\nonumber\\
&-&\left({r_q}^2+2 (x_0-1) x_0\right) \left({r_q}^2 \epsilon +2 x_0^4\right) \Bigg(2{r_q}^6 \epsilon ^2+{r_q}^4 x_0 \epsilon  \left(-8 x_0^3+2 x_0 \epsilon -3 \epsilon \right)\nonumber\\
&-&8 {r_q}^2 x_0^5 \left(x_0^3+2 (x_0-1) \epsilon \right)+4 (3-2x_0) x_0^9\Bigg)=0, 
\end{eqnarray}
where $\tilde{\Delta}_0$ is the value of $\tilde{\Delta}$ at $x=x_0$. When one solves Eq.~(\ref{xm}), one can get the value of the quantity $x_m$ as a function of $a$ , $r_q$ and $\epsilon$. The dependence of the spin $a$ on $x_m$ has been depicted in  Fig.~\ref{plot3} for a set of values of $\epsilon$  and $r_q$.    

\begin{figure}
	\begin{centering}
		\begin{tabular}{c c}
		    \includegraphics[scale=0.47]{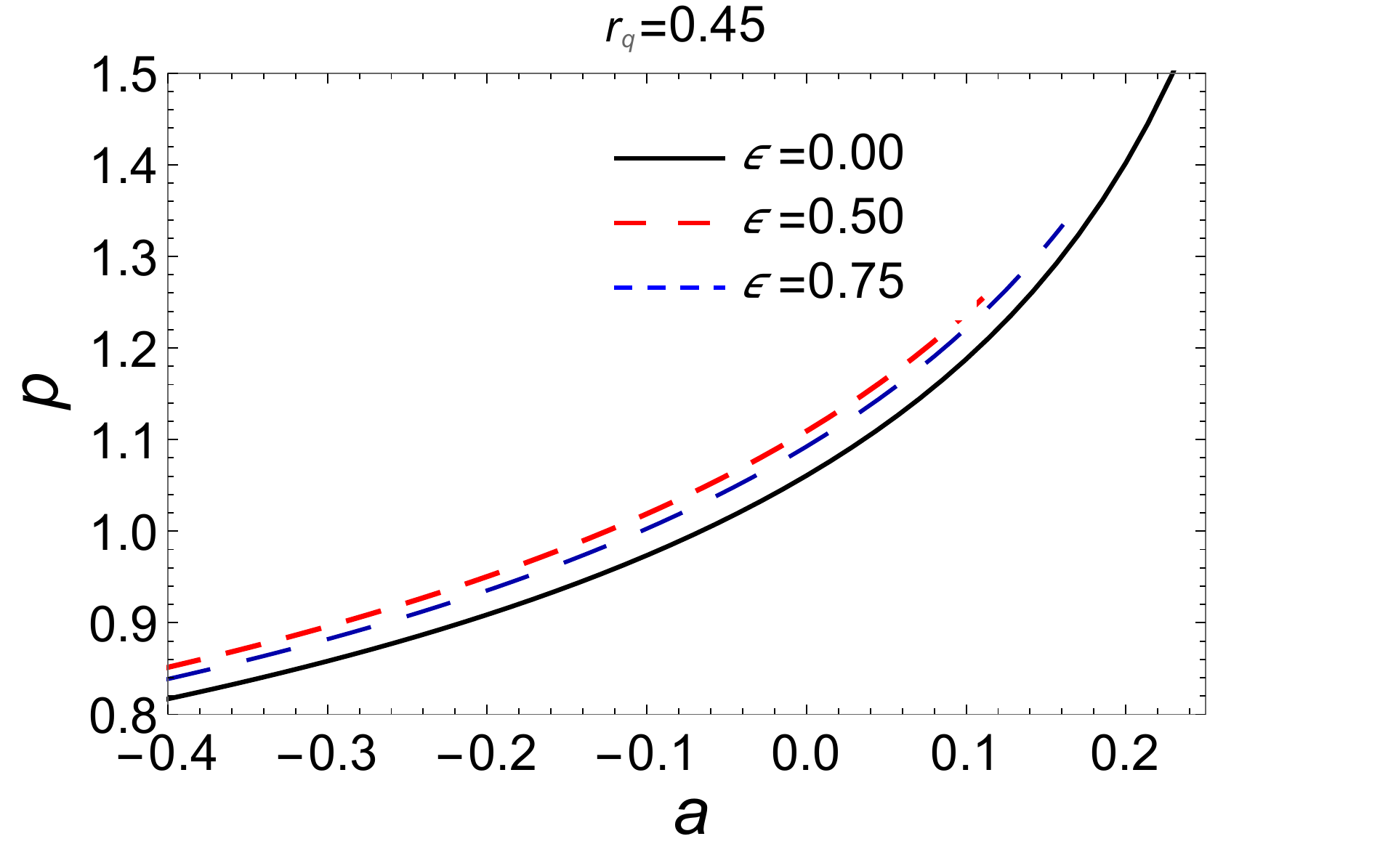}\hspace*{-0.9cm}&
		    \includegraphics[scale=0.47]{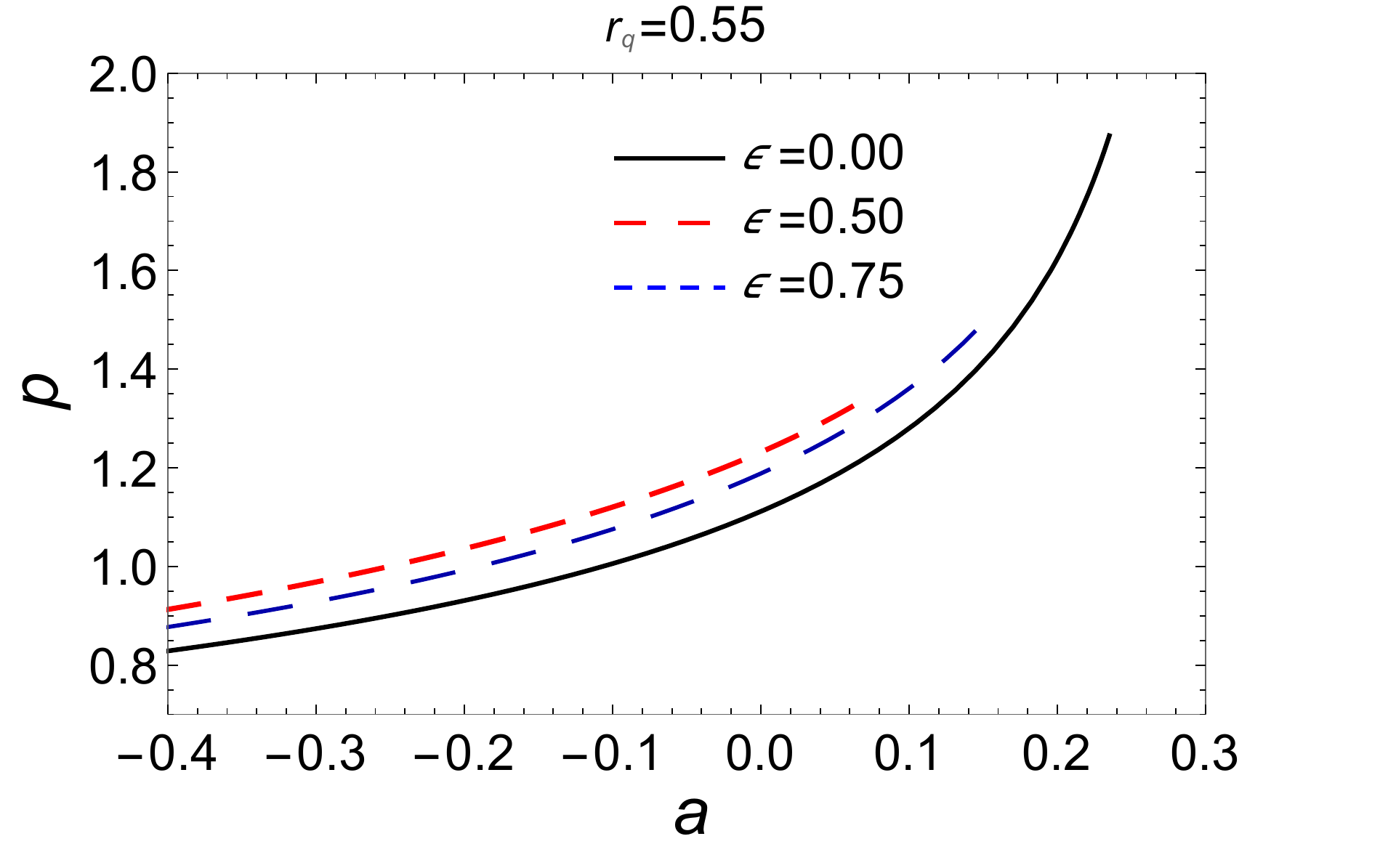}\\
			\includegraphics[scale=0.47]{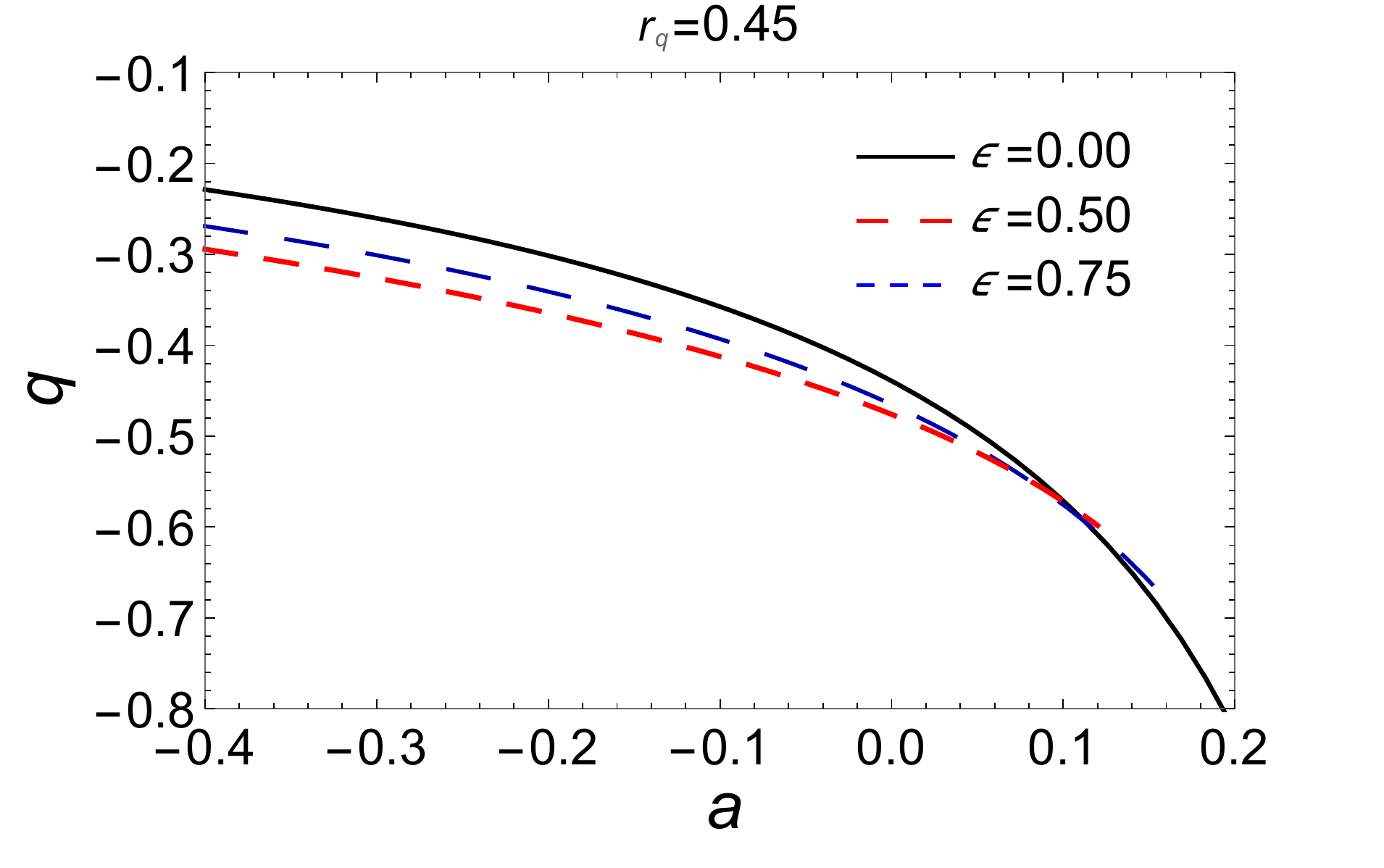}\hspace*{-0.9cm}&
		    \includegraphics[scale=0.47]{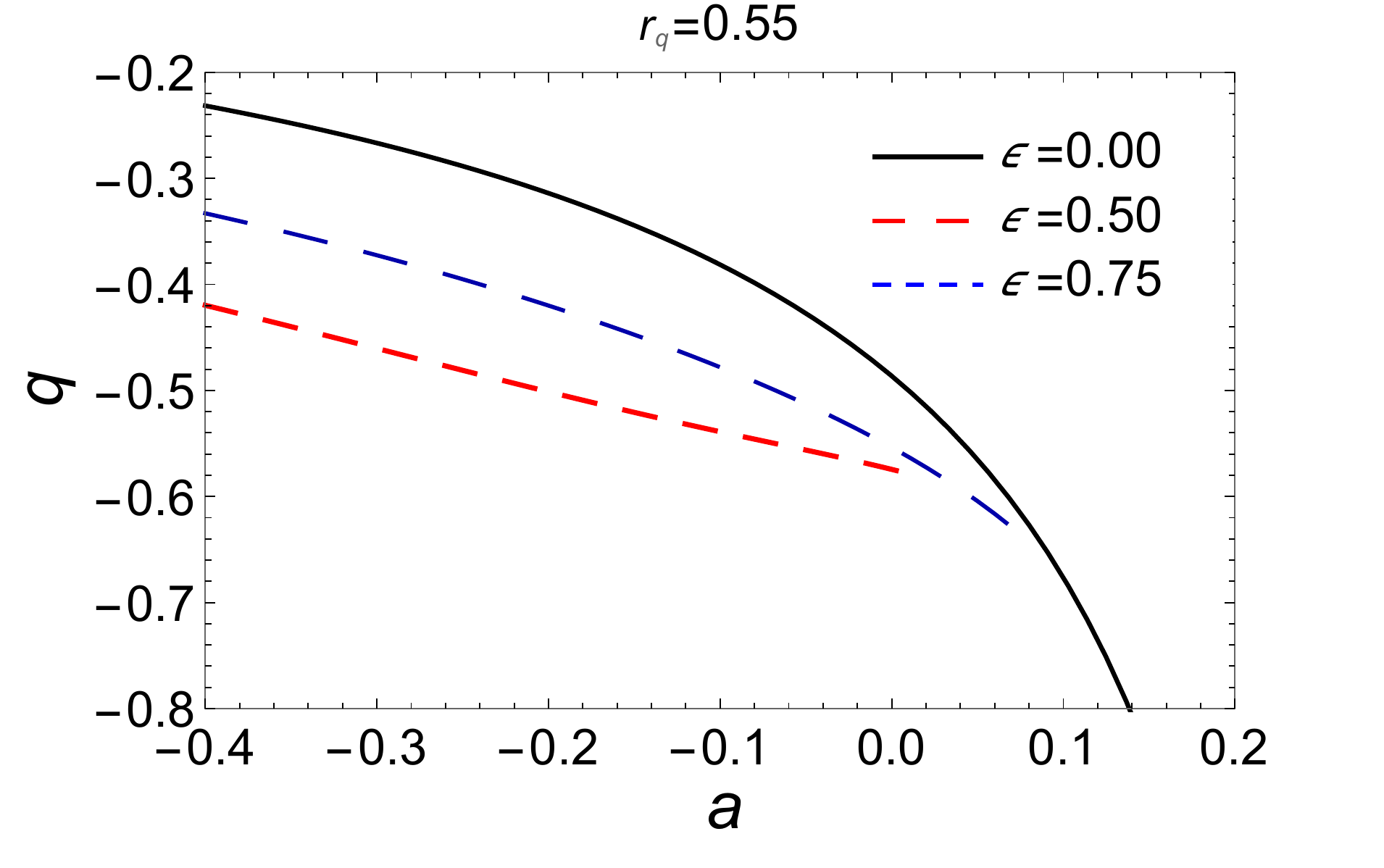}
			\end{tabular}
	\end{centering}
	\caption{The behaviour of the deflection coefficient $p$ vs the spin $a$ (upper panel) and the variation of the coefficient $q$ with spin $a$ (lower panel) for different values of $r_q$ and $\epsilon$.}\label{plot2}		
\end{figure}
\begin{figure*}[t]
	\begin{centering}
		\begin{tabular}{cc}
		 \includegraphics[scale=0.47]{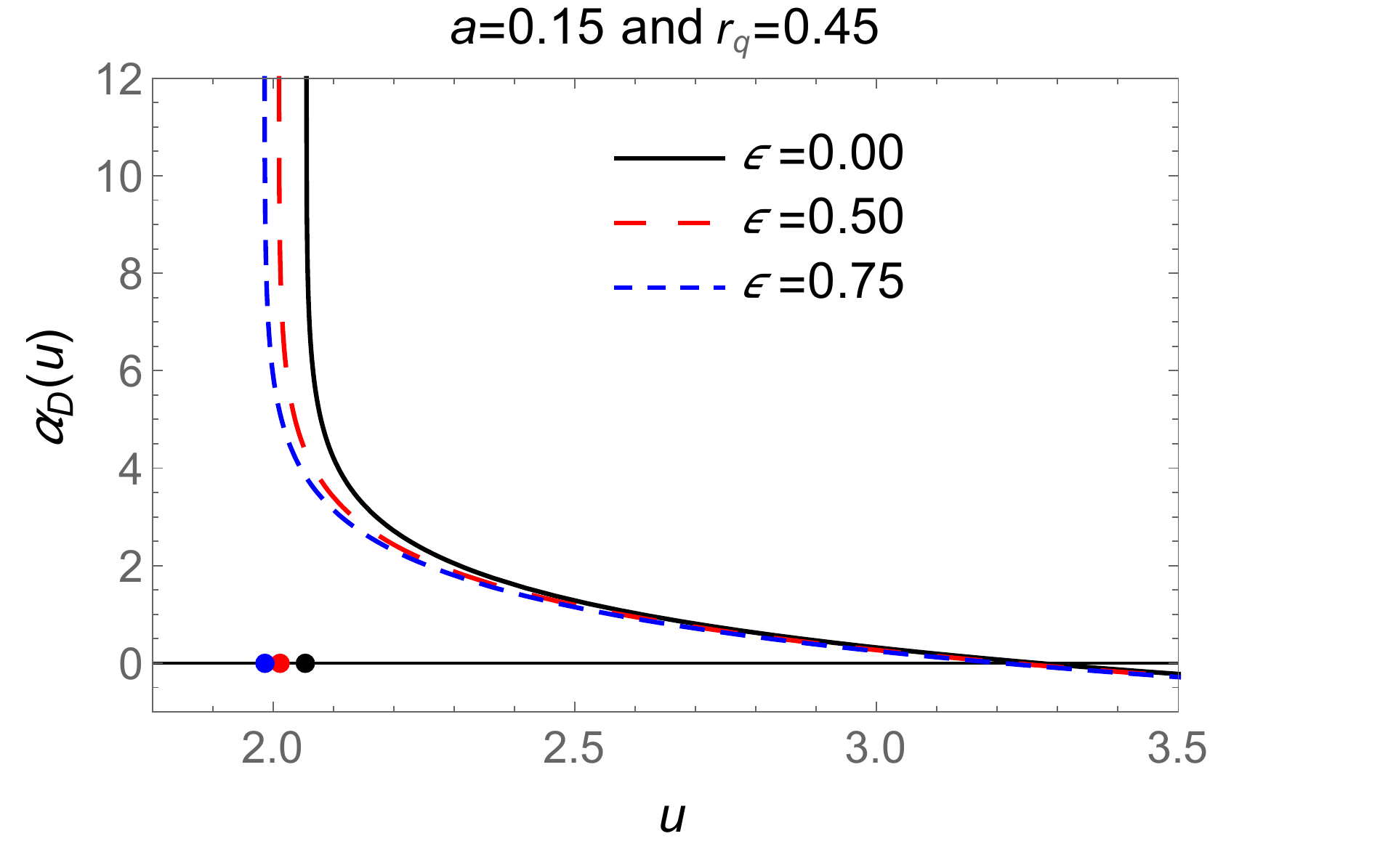}\hspace*{-0.9cm}&
		    \includegraphics[scale=0.47]{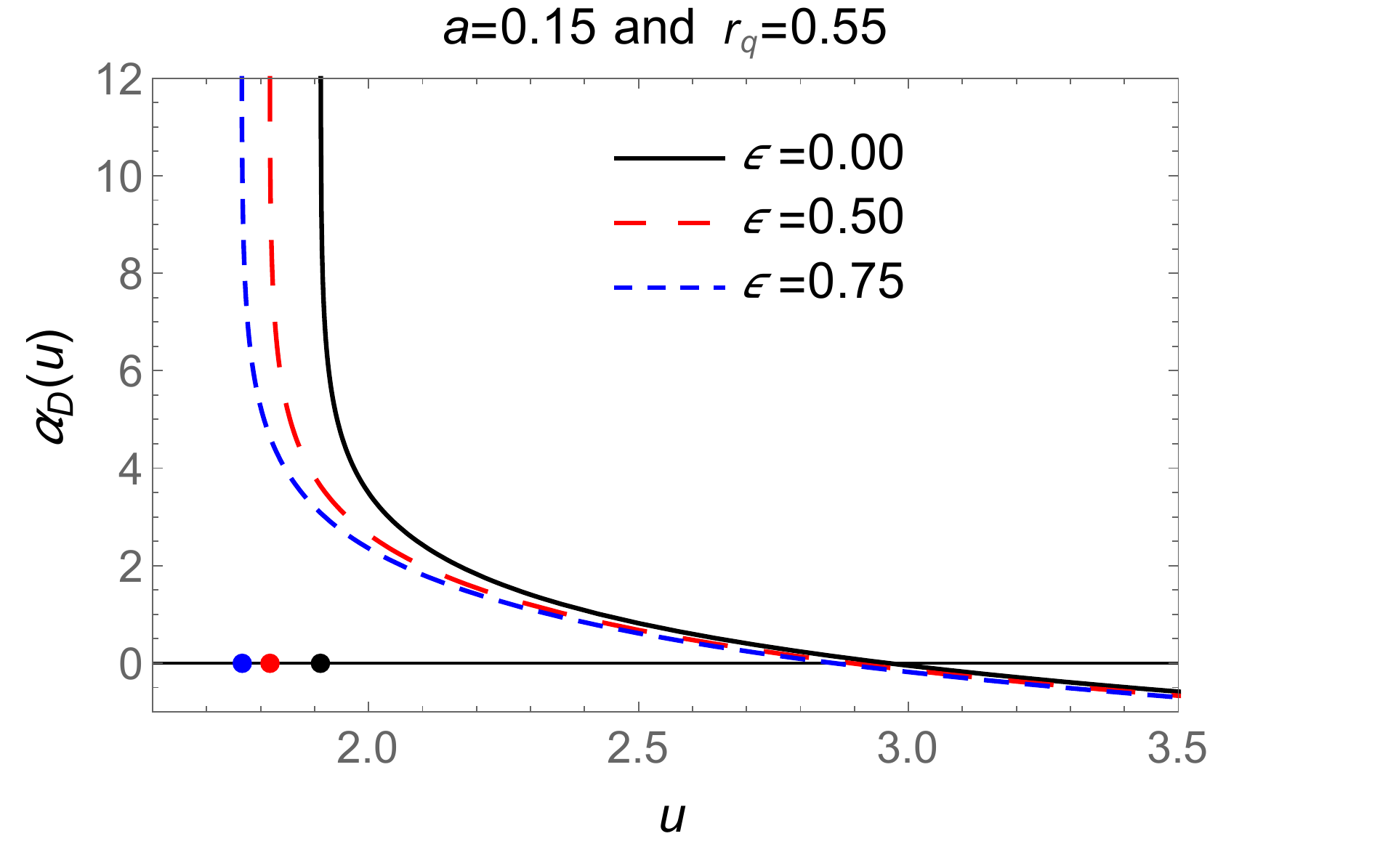}\\
		    \includegraphics[scale=0.47]{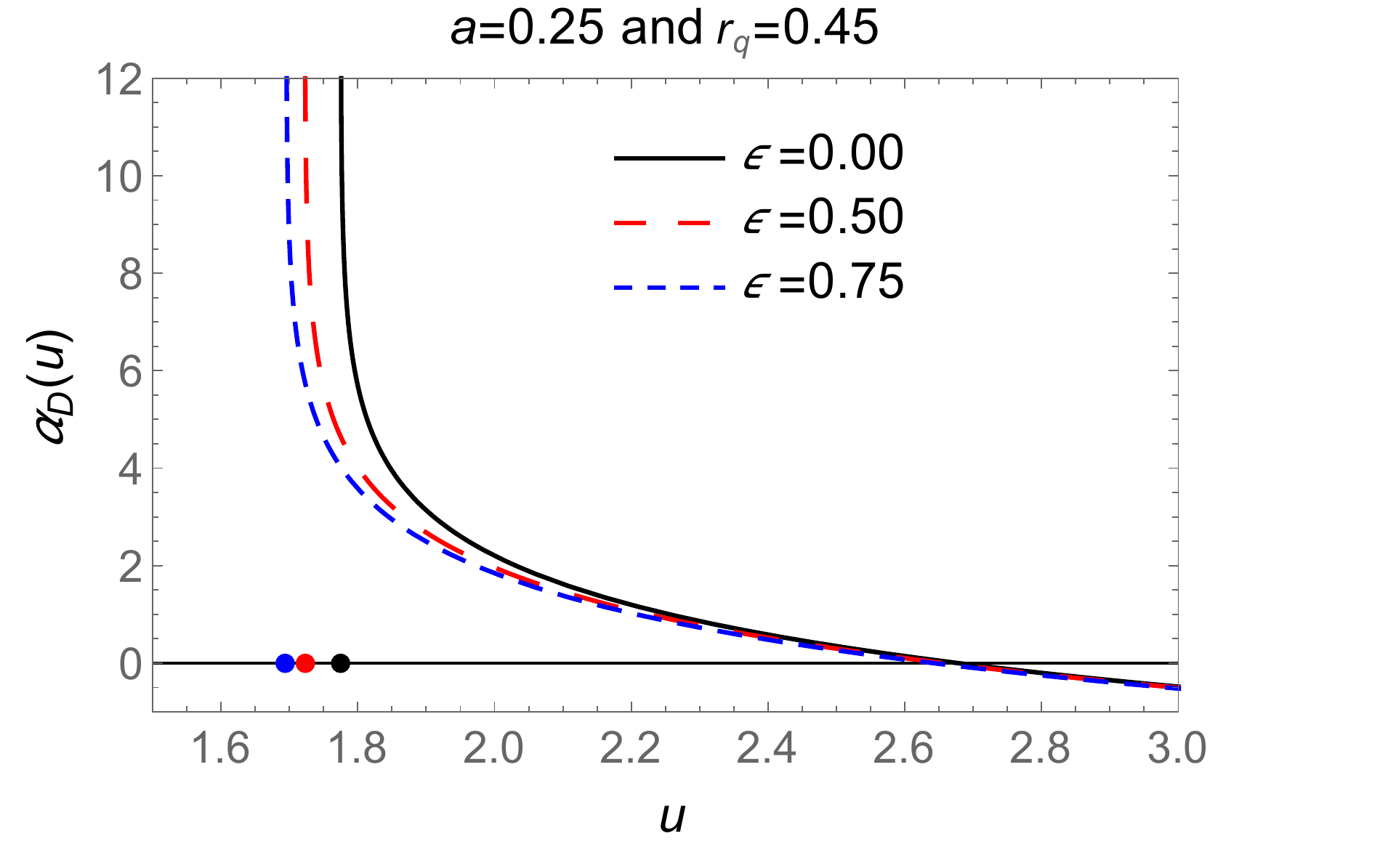}\hspace*{-0.9cm}&
		    \includegraphics[scale=0.47]{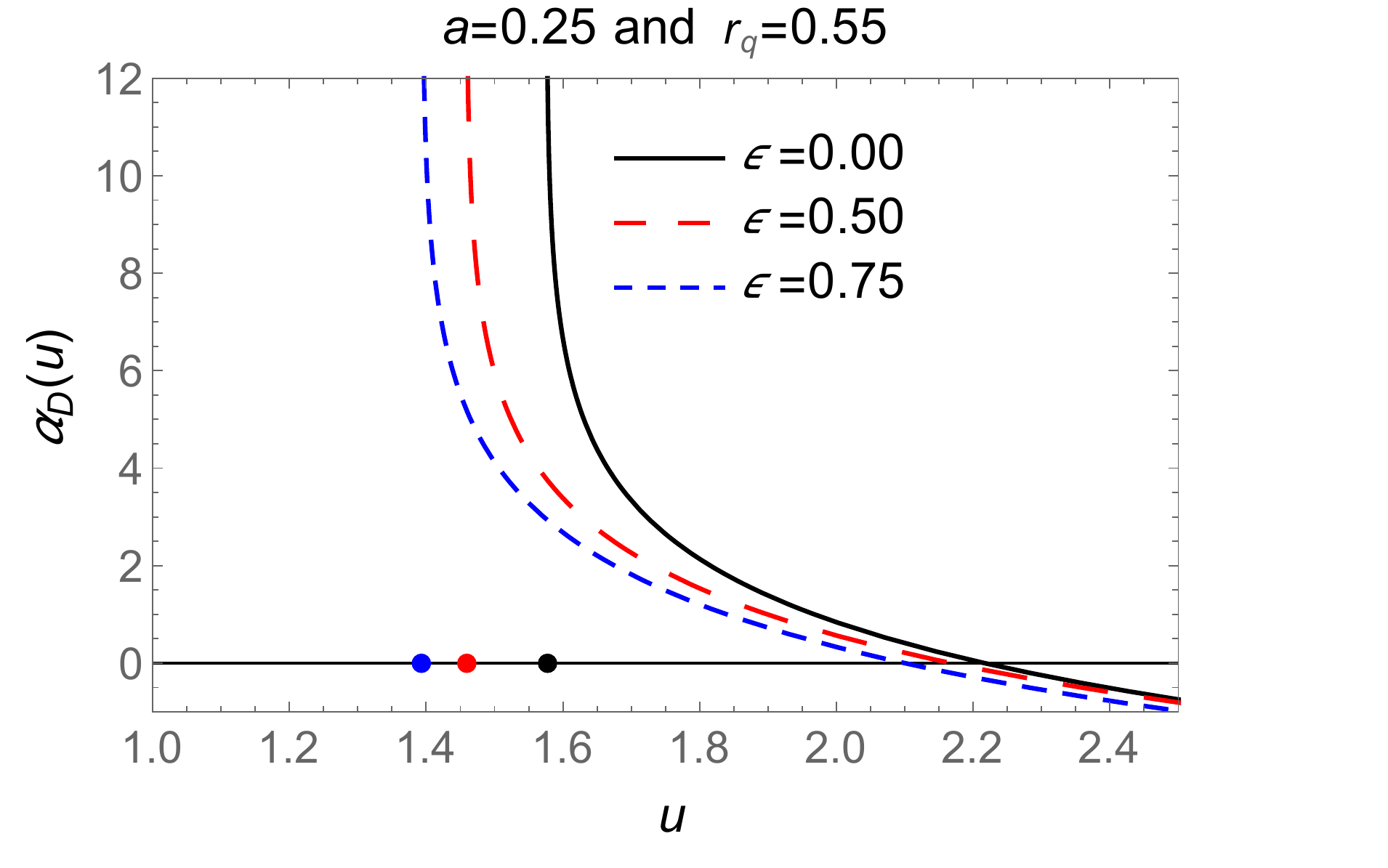}\\
			\end{tabular}
	\end{centering}
	\caption{Plot showing the behaviour of the light deflection angle for different values of $r_q$ and $\epsilon$. Points on the horizontal shows the divergence of the deflection angle at $u = u_m$}\label{plot4}		
\end{figure*}

In the limit $x_0 \to x_m $, the quantity $\alpha = 0$ and consequently we have $f(z,x_0) \approx 1/z $, thereby the integral (\ref{integral}) becomes infinitely large as $z \to 0$. We write the integral (\ref{integral}) as a combination of divergence and regular parts such that
\begin{eqnarray}
I(x_0)= I_D(x_0) + I_R(x_0),
\end{eqnarray}
with
\begin{eqnarray}\label{div}
I_D(x_0)&=&\int_{0}^{1} R(0,x_m)f_0(z,x_0)dz,\\
I_R(x_0)&=&\int_{0}^{1} [R(z,x_0)f(z,x_0)-R(0,x_0)f_0(z,x_0)]dz. \label{reg}
\end{eqnarray}
The integral Eq.~(\ref{div}) has an analytical solution
\begin{eqnarray}\label{phiR}
I_D(x_0)&=& \frac{2 R(0,x_m)}{\sqrt{\gamma}} \log\Bigg(\frac{\sqrt{\gamma + \alpha}+\sqrt{\gamma}}{\sqrt{\alpha}}\Bigg),
\end{eqnarray}
As $\alpha=0$ at $x_0=x_m$, the right hand side of Eq.~(\ref{div}) has a infinity at $x_0=x_m$ as can be seen from the expression inside the logarithm above. Therefore the regular part in contained in the integral of Eq.~(\ref{reg}). Since it has significant contribution  upto order of $(x_0-x_m)$, we can take the regular parts as
\begin{eqnarray}\label{phiT}
I_R(x_m) = \int_{0}^{1} [R(z,x_m)f(z,x_m)-R(0,x_m)f_0(z,x_m)]dz,
\end{eqnarray} 
whose behavior can be seen from the numerical plots, Fig.~\ref{plot4}. Now using Eqs.~(\ref{phiR}) and (\ref{phiT}), we express the quantity $\alpha_D$ as follows
\begin{eqnarray}\label{def}
\alpha_{D}(\theta)=-p \log\Big(\frac{\theta D_{OL}}{u_m}-1\Big)+q + \mathcal{O}\left(u-u_m\right),
\end{eqnarray}
where the quantities $p$ and $q$ in Eq.(\ref{def}) for the strong gravitational field limit are cast as
\begin{eqnarray}
p = \frac{R(0,x_m)}{2\sqrt{\gamma _m}}, ~~~ \textrm{and}~~~ q = -\pi +I_R(x_m) + p\log\frac{c x_m^2 }{u_m^2}
\end{eqnarray}
which is a polynomial of various parameters and is expressed as
\begin{small}
\begin{eqnarray}\label{abar}
p = \frac{2 \sqrt{\left({r_q}^2 \epsilon +2{x_0}^4\right) \left(2 {r_q}^2 \left(2 a^2 \epsilon +{x_0}^4+({x_0}-1) {x_0} \epsilon \right)+{r_q}^4 \epsilon +4 ({x_0}-1) {x_0}^5\right)}}{\sqrt{2 a^2+{r_q}^2+2 ({x_0}-1){x_0}} \sqrt{c_1\gamma_m}},
\end{eqnarray}
\end{small}
where
\begin{eqnarray}
c_1 &=&{2{x_0}^4 \left(a^2 \left(2 {x_0} ({x_0}+1)-{r_q}^2\right)+2{x_0}^4\right)-{r_q}^2 \epsilon  \left(4 a^4+a^2 \left({r_q}^2+2 {x_0} (3{x_0}-1)\right)+2 {x_0}^4\right)}.
\end{eqnarray}
Series expanding the Eq.~(\ref{angmom}) in terms of $(x_0-x_m)$, we have
\begin{eqnarray}
u-u_m &\approx& c_2(x_0-x_m)^2.
\end{eqnarray}
The analytical expression for $c_2$ is very large and we do not write it here. We use the expression of $c_2$ in the numerical investigations of the light deflection coefficients $p$ and $q$. Remember that all the expressions with the subscript $m$ are obtained at $x_0=x_m$. The quantities $p$ and $q$, in Eq.~(\ref{abar}), appear in the calculation of the total azimuthal angle and they are called the deflection coefficients in the strong field regime. We plot them in Fig.~\ref{plot2}, which depict that $p$ and $q$, show the opposite behavior with the different values of spin parameter $a$. As expected these quantities become infinitely large as we increase the values of $a$, thereby indicating the validity of the coefficients at higher rotation parameter cease to exist. As a limiting cases those results of strong-field deflection coefficients reduce to the corresponding limits of Kerr-Newman black holes when $\epsilon\to 0$, the Kerr black holes when $r_q\to 0$ and also the Schwarzschild black holes when $a\to 0$, $r_q\to 0$, and $\epsilon\to 0$.

\section{Observables and relativistic images}\label{Sec4}
We adopt to decribe the strong gravitational lensing, the lens equations. There exist several methods to describe the lens equations as they principally dependent on different choices of the variables. In describing the gravitational lensing we place the black hole at the origin such that at one side there is the observer and at another side there is the light source. The light rays coming from the illuminating source (S) gets deviated from its original path while passing the black hole (L) due to the curvature and ultimately reach to the observer (O). The line connecting the black holes and the observer and the image that the observer sees is an optical axis OL and it will be deviated at an angle $\theta$ with resepct to OL. Similarly, the light source will be aligned at $\beta$ angle with OL. The emitted light rays makes an angle $\alpha_{D}(\theta)$ when detected by the observer.\\
As mentioned earlier there are various mathematical formulations to interpret the lensing phenomena. Among them Ohanian lens equation is the best approximation \citep{Ohanian:1987pc} to describe the positions of observer and the source as
\begin{eqnarray}\label{oho}
\xi &=& \frac{D_{OL}+D_{LS}}{D_{LS}}\theta-\alpha_{D}(\theta),   
\end{eqnarray} 
where the angle $\xi\in[-\pi,\pi]$ connects the optical axis and the source directions. $D_{OL}$ is lens to the observer distance, while $D_{LS}$ is to that of lens to source distance. The angles $\xi$ and $\beta$ are found to follow the relation \cite{Ohanian:1987pc, Ghosh:2020spb}
\begin{eqnarray}\label{rel}
\frac{D_{OL}}{\sin(\xi-\beta)} &=& \frac{D_{LS}}{\sin \beta}
\end{eqnarray}
For the completeness of the above relations we choose $\theta$, $\xi$, and $\beta$  to be tinier because in this case, the relativistic images formed by the black holes are prominent. The light rays coming from source S and make many loops while encountering the black holes, so that the deflection angle $\alpha $ is replaced by $2n\pi + \Delta\alpha _n$, where the integer $n \in N $ represents the number that counts the loops and $ 0<\Delta\alpha _n \ll 1$. The Eq.~(\ref{oho}) together with Eq.~(\ref{rel}) for smaller values of $\theta$, is rewritten as
\begin{eqnarray}\label{lensequation}
\beta &=& \theta -\frac{D_{LS}}{D_{OL}+D_{LS}} \Delta\alpha _n.
\end{eqnarray}  
Next, Eq.~(\ref{lensequation}) is utilized to extract the information regarding the image positions. For a critical impact parameter $u_m$, which is a function of the distance of the photon orbit radius $x_m$, $\alpha_D(\theta)$ becomes infinitely large. Each loop of the light rays near event horizon of the black holes, there exits a certain value of $u$ at which photons reach from source to the observer. Therefore, on both sides of the black holes an infinite number of  relativistic images are constructed. Now equation~(\ref{def}) with $\alpha_{D}(\theta_n{^0}) = 2n\pi $ reads as
\begin{eqnarray}
\label{theta}
\theta_n{^0} &=& \frac{u_m}{D_{OL}}(1+e_n),
\end{eqnarray}  
where 
\begin{eqnarray}
e_n &=& e^{\frac{q-2n\pi}{p}}.
\end{eqnarray}

The Taylor expansion of the deflection angle $\alpha_{D}(\theta)$ around $\theta_n{^0}$ to the first order in $(\theta-\theta_n{^0})$ reads \cite{Ghosh:2020spb}
\begin{eqnarray}
\alpha_{D}(\theta) &=& \alpha_{D}(\theta_n {^0}) +\frac{\partial \alpha_{D}(\theta)}{\partial \theta } \Bigg |_{\theta_n{^0}}(\theta-\theta_n{^0})+\mathcal{O}(\theta-\theta_n{^0}). 
\end{eqnarray}
On utilizing Eq.(\ref{theta}) and defining $\Delta\theta_n= (\theta-\theta_n{^0}) $ one gets
\begin{eqnarray}
\Delta\alpha_n &=& -\frac{pD_{OL}}{u_m e_n}\Delta\theta_n.
\end{eqnarray}
Now the final expression for lens equation (\ref{lensequation}) reads \cite{Ghosh:2020spb}
\begin{eqnarray}
\label{final}
\beta &=& ( \theta_n{^0} + \Delta\theta_n )+\frac{D_{LS}}{D_{OL}+D_{LS}}\Bigg(\frac{p D_{OL}}{u_m e_n}\Delta\theta_n\Bigg).
\end{eqnarray}
Substituting the value of $\Delta\theta_n= (\theta-\theta_n{^0}) $ and then ignoring the second term in Eq. (\ref{final}) as it contributes very less as compared to the second term, we have 
\begin{eqnarray}\label{angpos}
\theta_n &=& \theta_n{^0} + \frac{D_{OL}+D_{LS}}{D_{LS}}\frac{u_me_n}{D_{OL}p}(\beta-\theta_n{^0}).
\end{eqnarray}
\begin{figure*}[t]
	\begin{centering}
		\begin{tabular}{cc}
		    \includegraphics[scale=0.47]{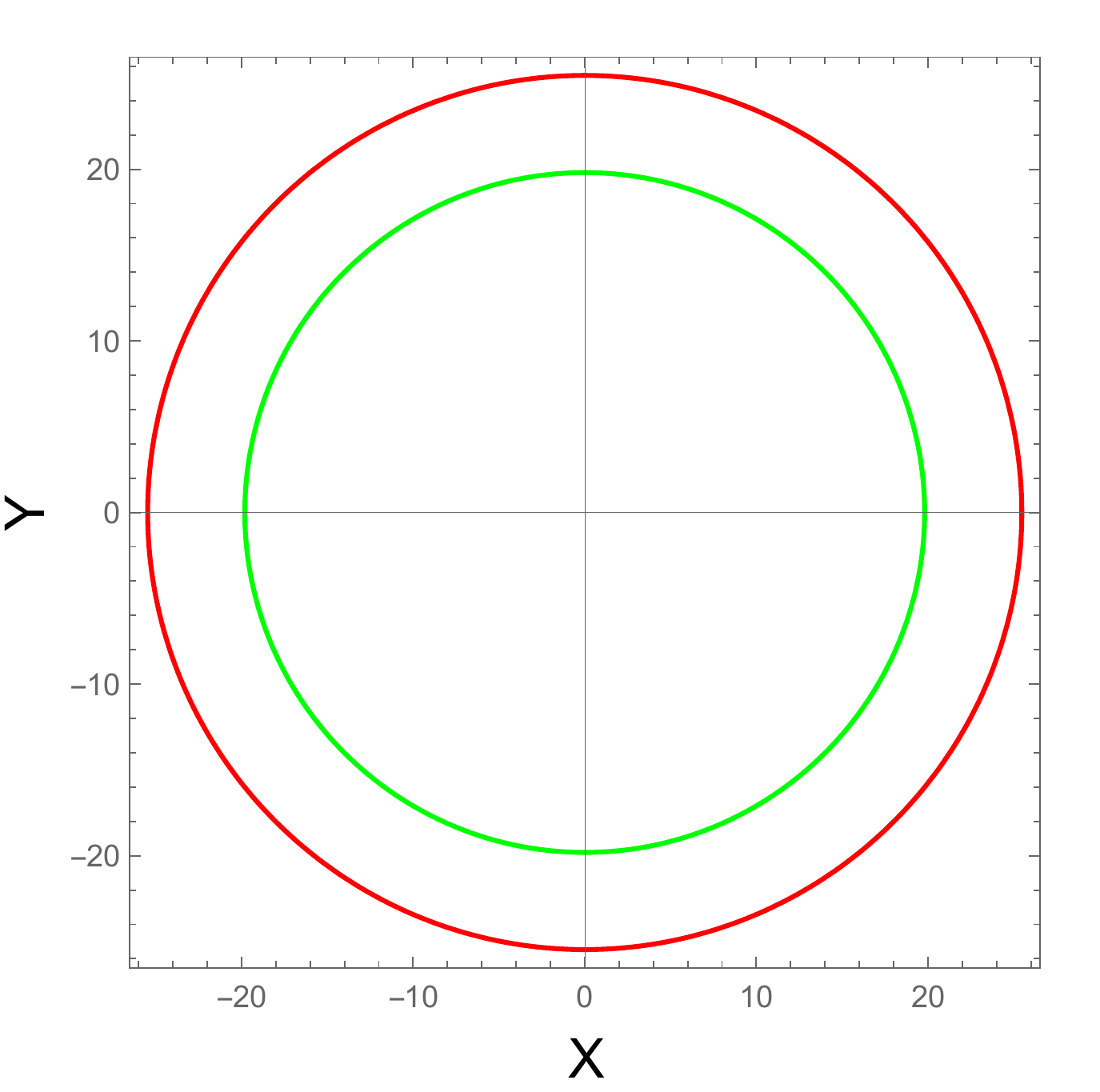}
		    \end{tabular}
	\end{centering}
\caption{Plots for the outermost Einstein rings for black holes at the center of nearby galaxies in the framework  of Schwarzschild geometry. Red line corresponds to the $SgrA^*$ and green line to that of $M87$ \cite{Ghosh:2020spb}}
\label{plotER}	 
\end{figure*}	
Next we discuss the most striking features of the gravitational lensing, the formation of the Einstein's ring. Einsten's rings are formed when we have the point lens perfectly aligned in the line sight of the source such that the light from the source are spreading in all directions equally. The complex lens systems may lead to the formation of multiple images \cite{Ohanian:1987pc,Virbhadra:1999nm,Virbhadra:2002ju}, see for double Einstein's ring formation \cite{Gavazzi:2008} due to two different sources. The  relativistic Einstein's rings are formed when the deflection angle , $\alpha\geq 2\pi$. For the lens and the observer to be perfectly oriented ($\beta=0$), and the lens are situated perfectly at the center of the observer and the source, then the equation Eq. (\ref{angpos}) reads \cite{Ghosh:2020spb}
	\begin{eqnarray}\label{Ering2}
		\theta_n^{E} &=& \left(1-\frac{2 u_me_n}{D_{OL}p} \right) \left(\frac{u_m}{D_{OL}}(1+e_n)\right).
	\end{eqnarray} 
For $D_{OL} \gg u_m$, the angular radius (\label{Ering2}) for Einstein's ring reduces to 
	\begin{eqnarray}\label{Ering3}
		\theta_n^{E} &=& \frac{u_m}{D_{OL}}\left(1+e_n \right).
	\end{eqnarray} 

\begin{figure*}[t]
	\begin{centering}
		\begin{tabular}{c c}
		    \includegraphics[scale=0.50]{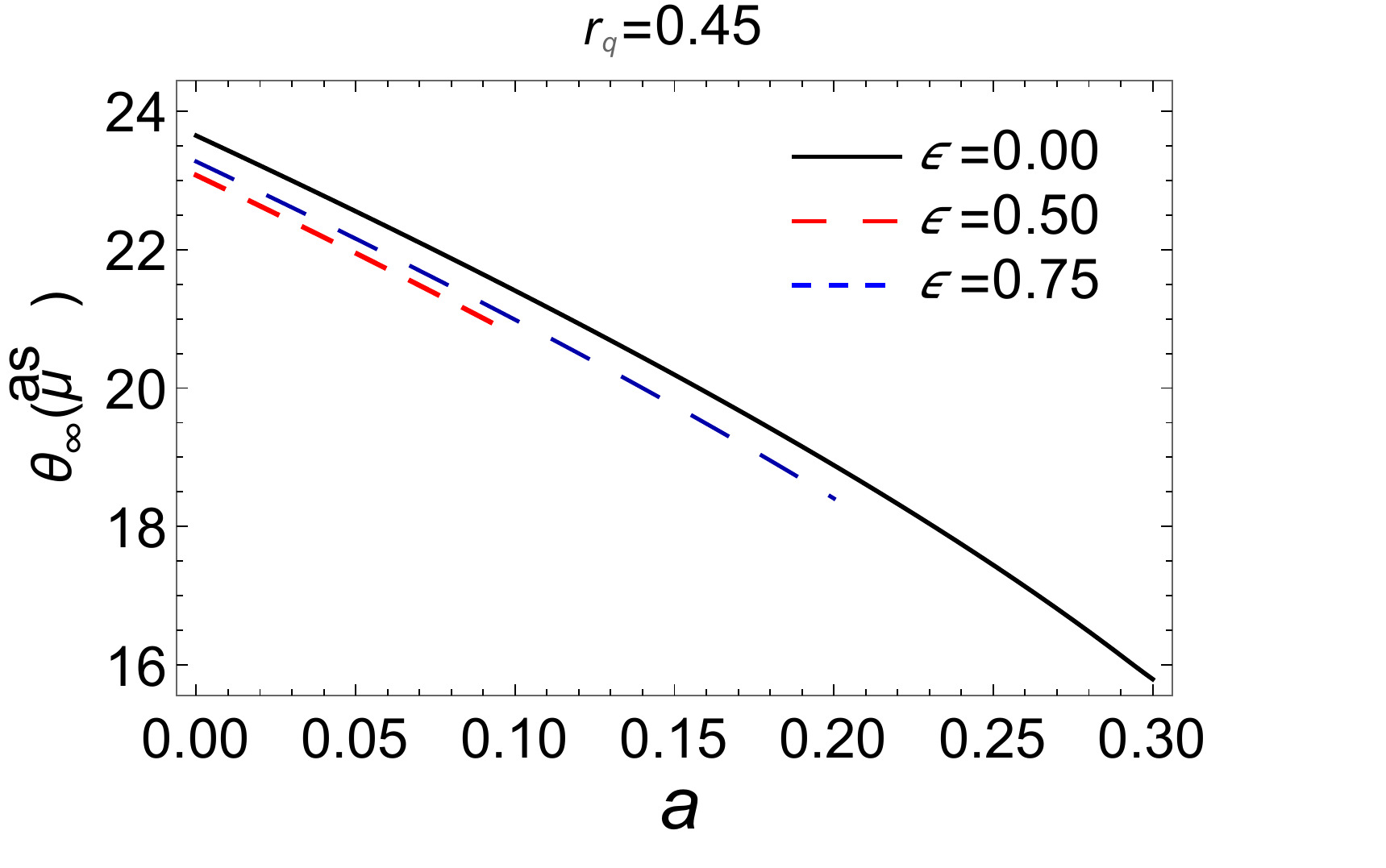}\hspace*{-0.7cm}
		    \includegraphics[scale=0.50]{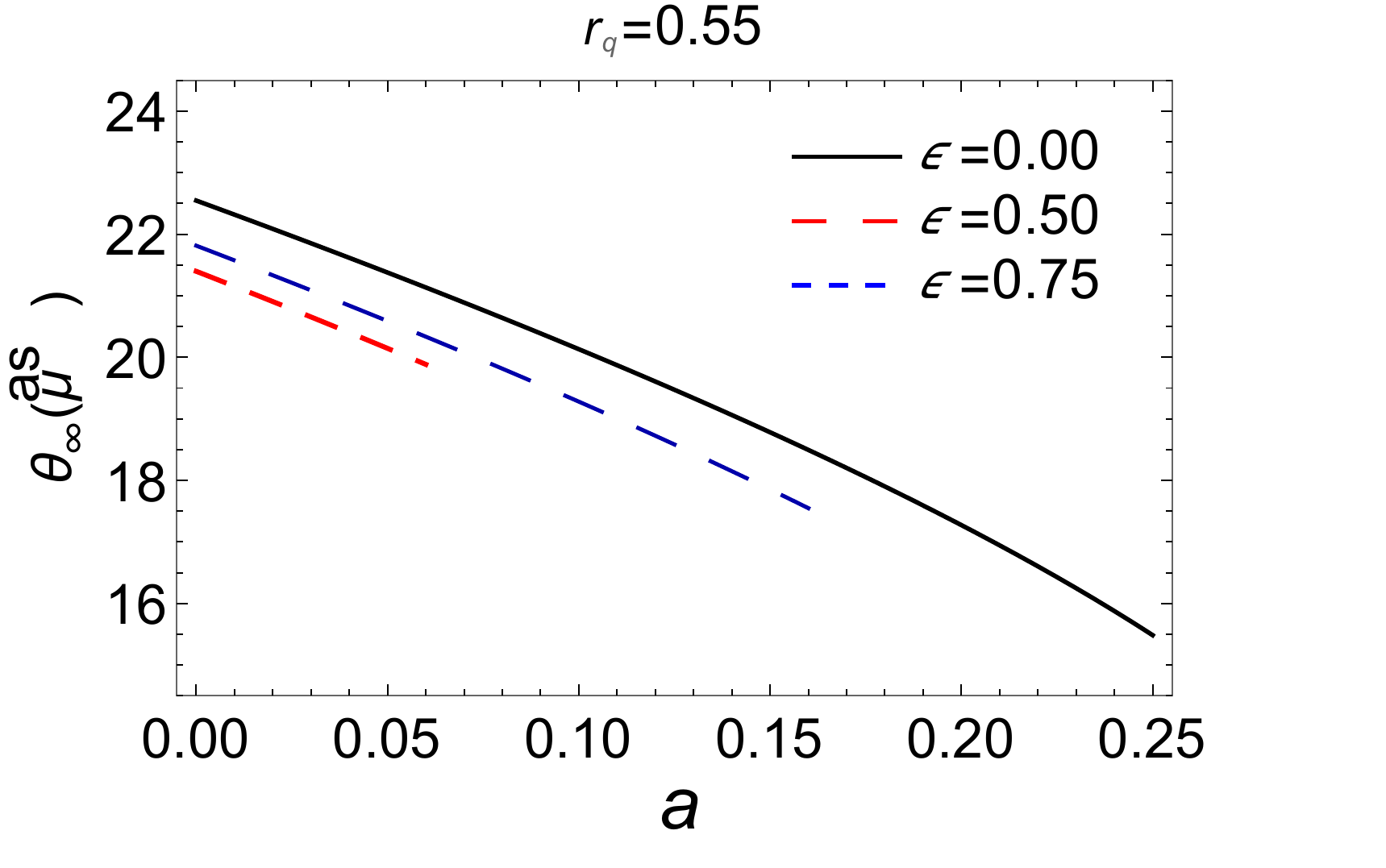}\\
			\includegraphics[scale=0.50]{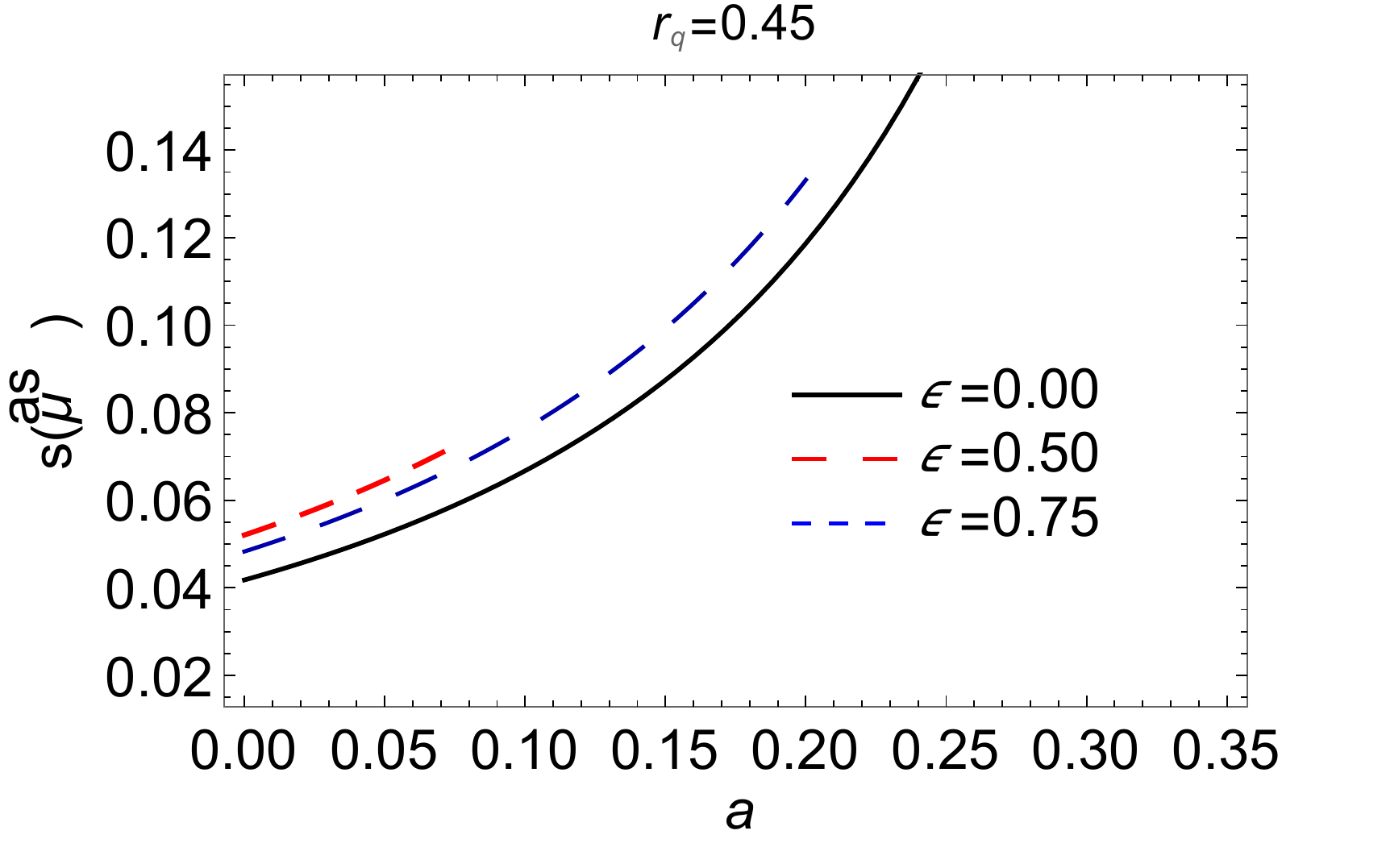}\hspace*{-0.7cm}
		    \includegraphics[scale=0.50]{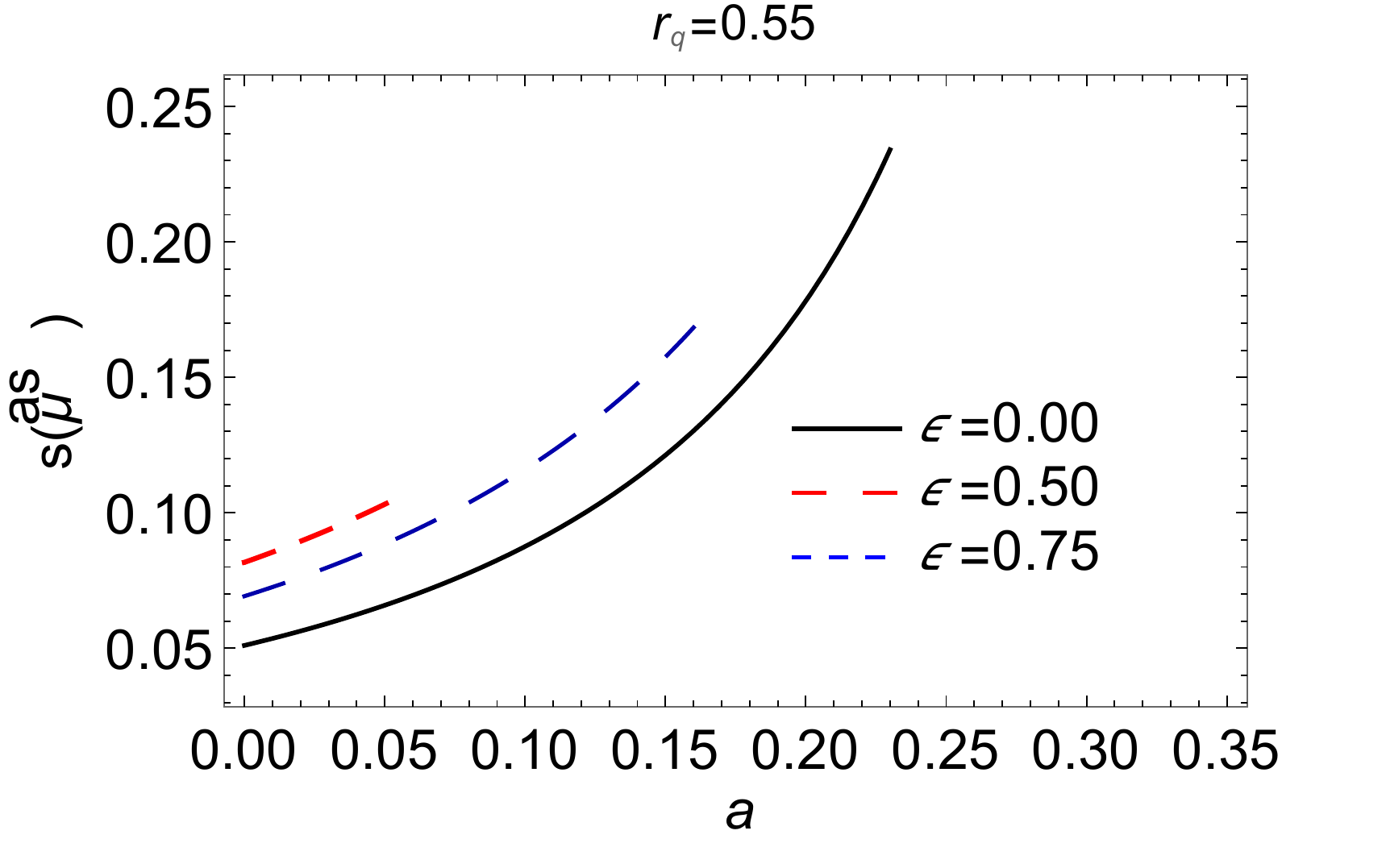}\\
			\includegraphics[scale=0.50]{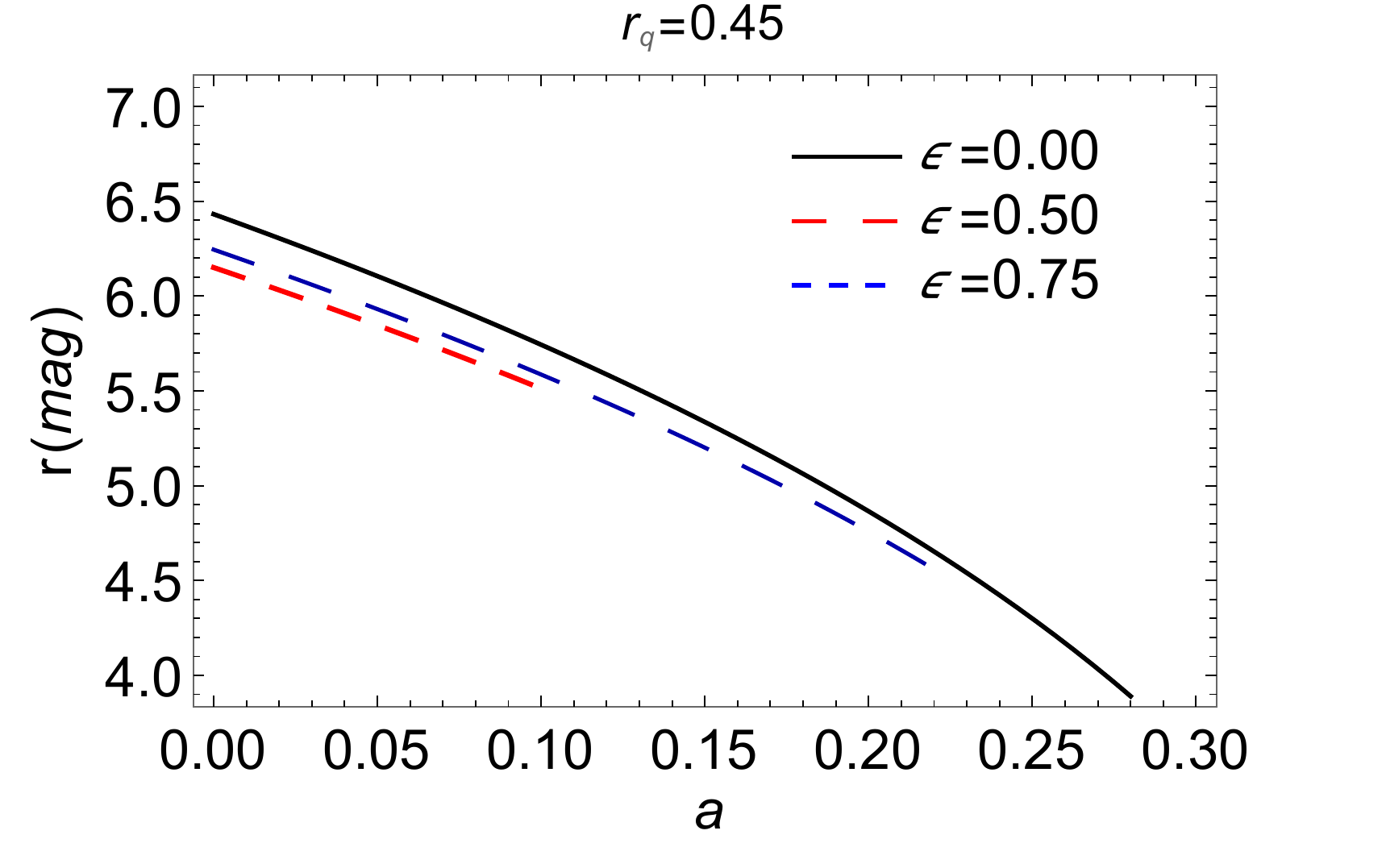}\hspace*{-0.7cm}
		    \includegraphics[scale=0.50]{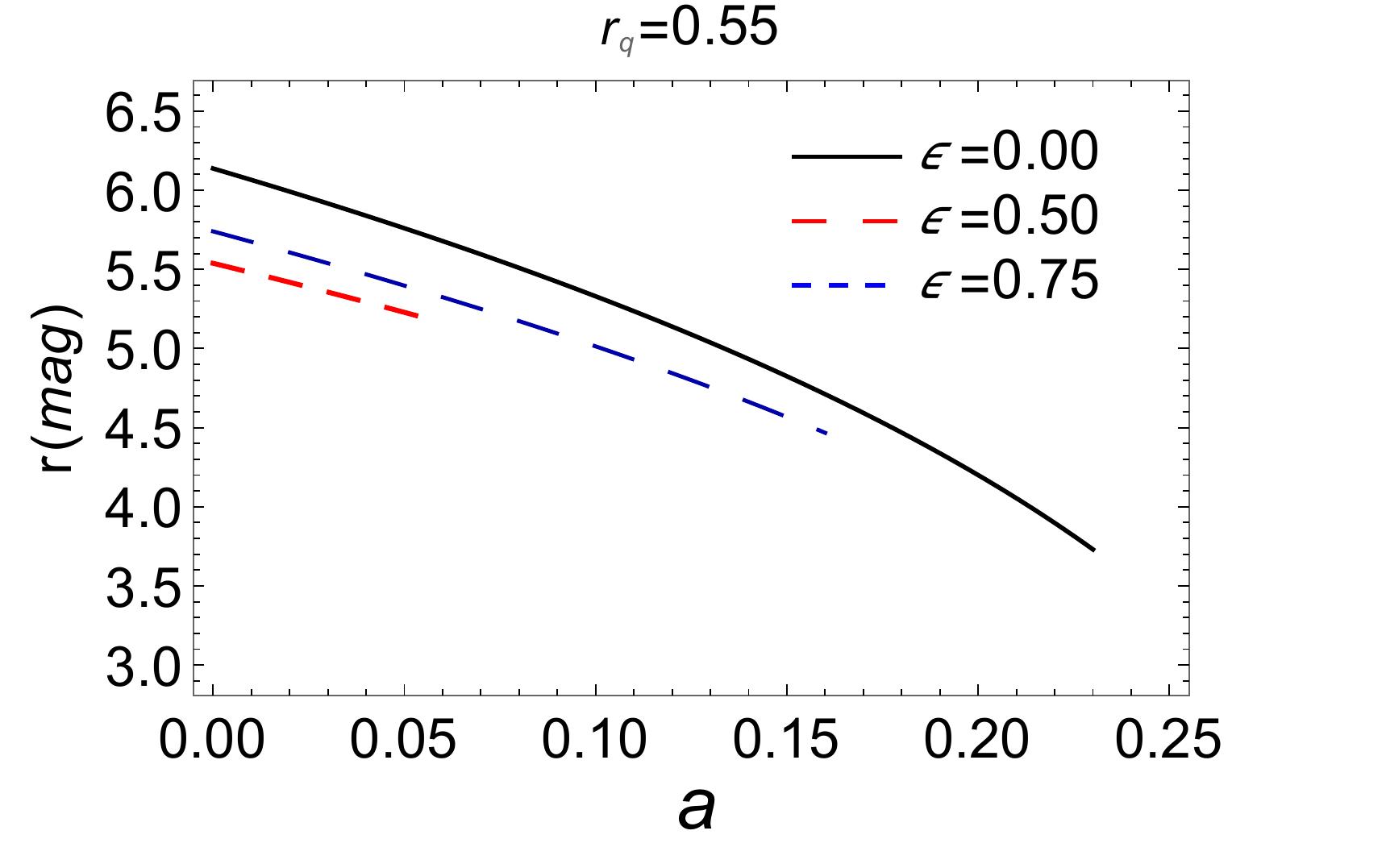}
			\end{tabular}
	\end{centering}
	\caption{Plot showing the variation of lensing observables  $\theta_{\infty}$, $s$, and $r_{mag}$ as a function of $a$ for different values of $r_q$ and $\epsilon$ for SgrA*.}\label{plot1a}		
\end{figure*}
\begin{figure*}[t]
	\begin{centering}
		\begin{tabular}{c c}
		    \includegraphics[scale=0.50]{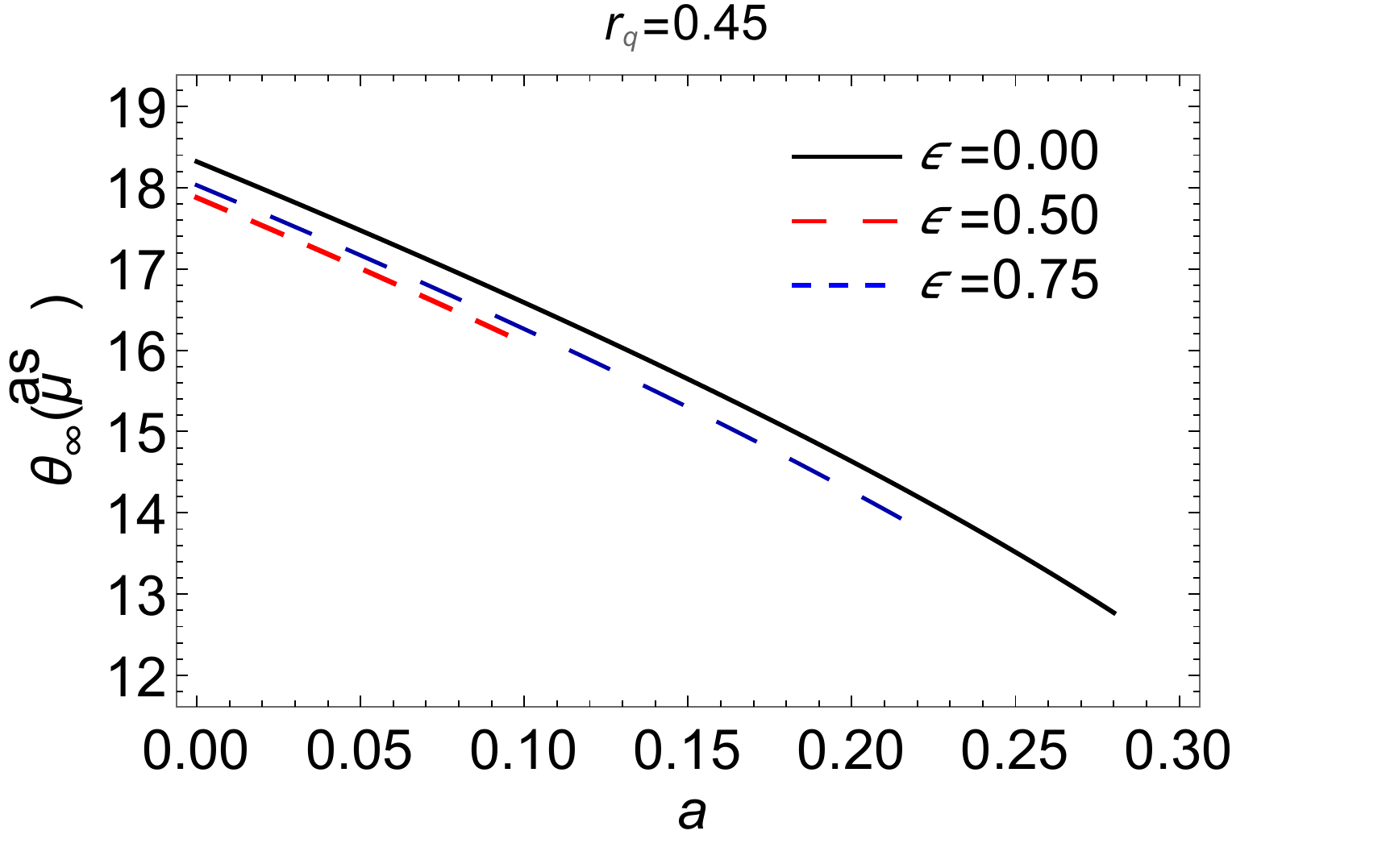}\hspace*{-0.7cm}
		    \includegraphics[scale=0.50]{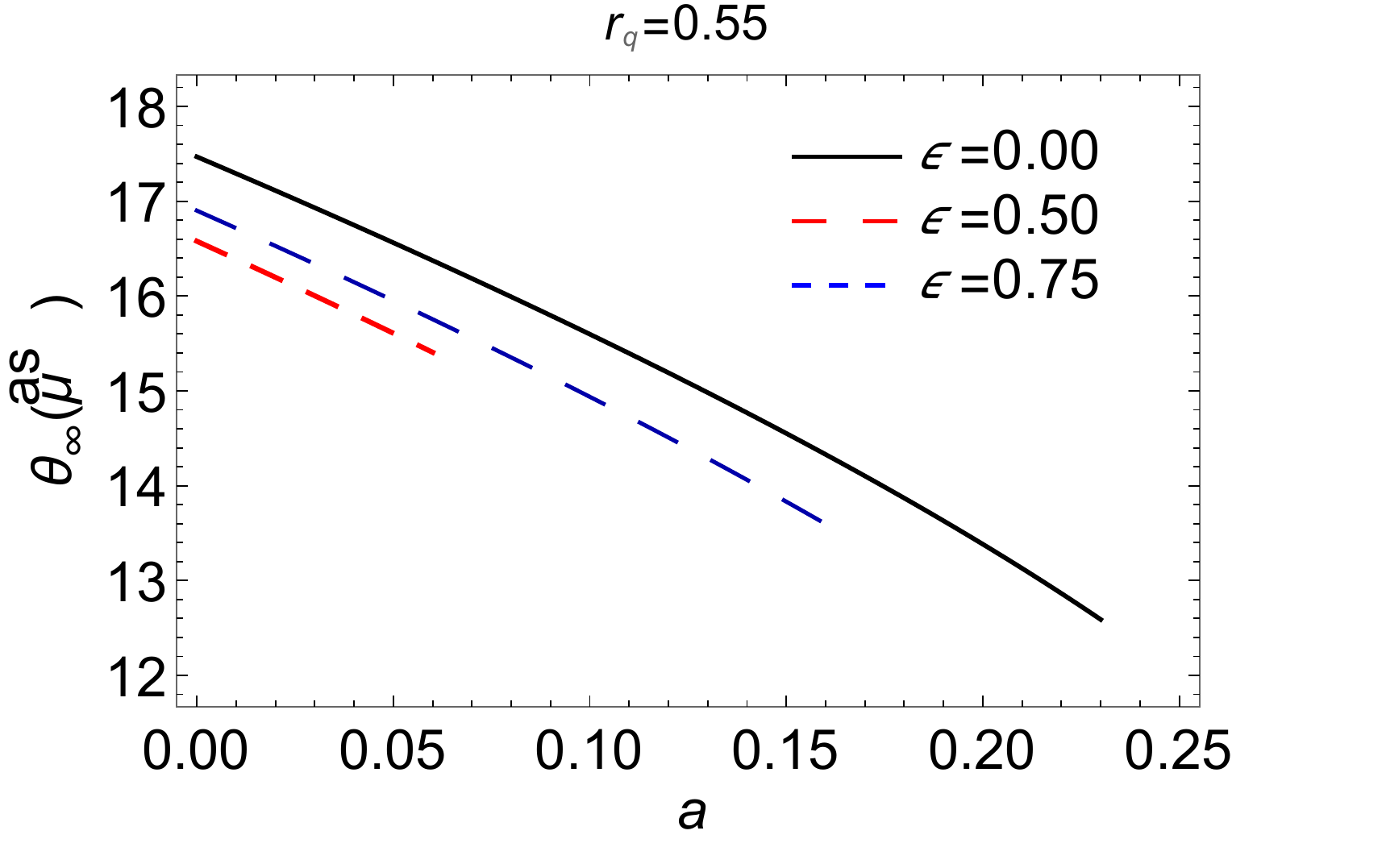}\\
			
			\includegraphics[scale=0.50]{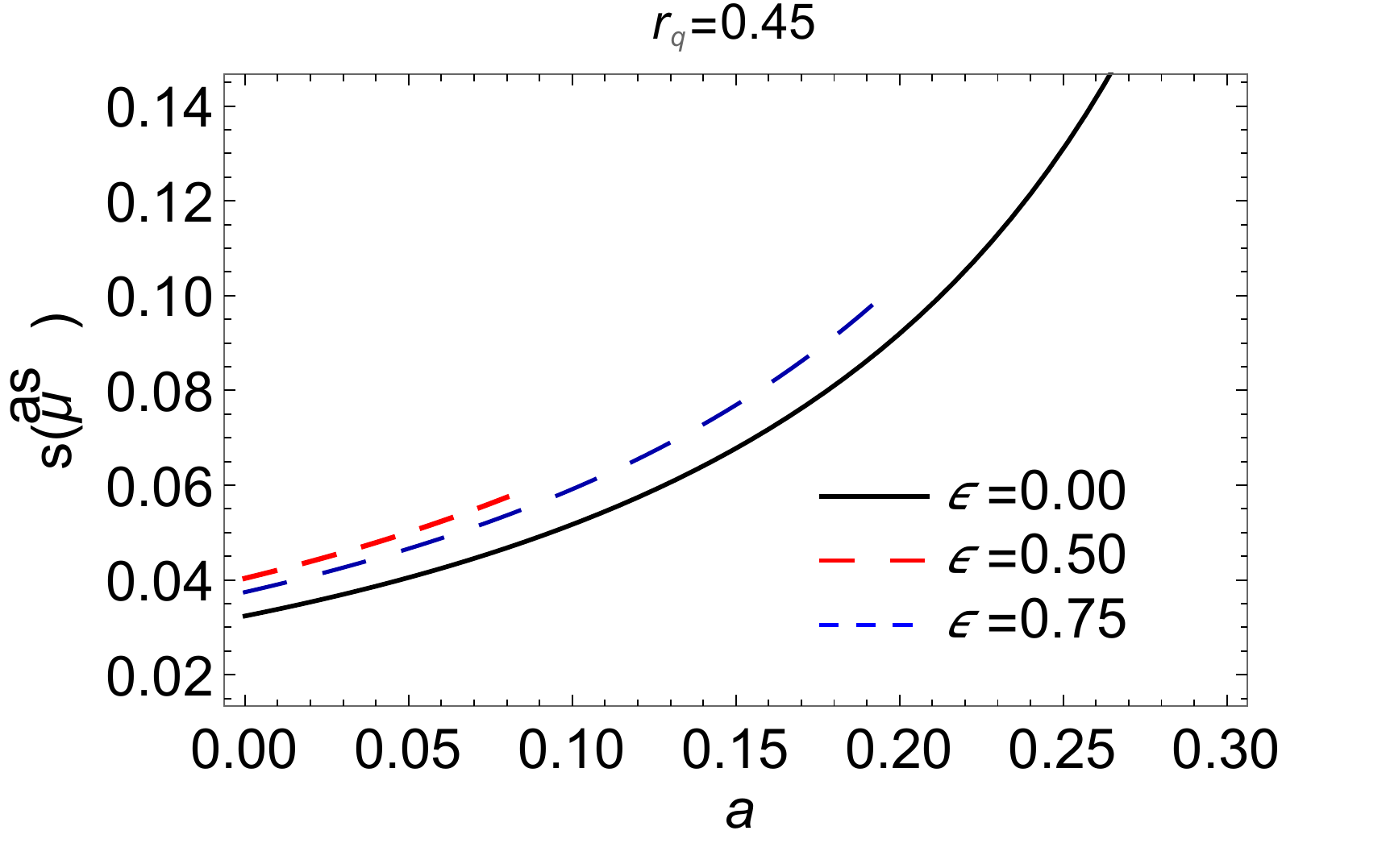}\hspace*{-0.7cm}
		    \includegraphics[scale=0.50]{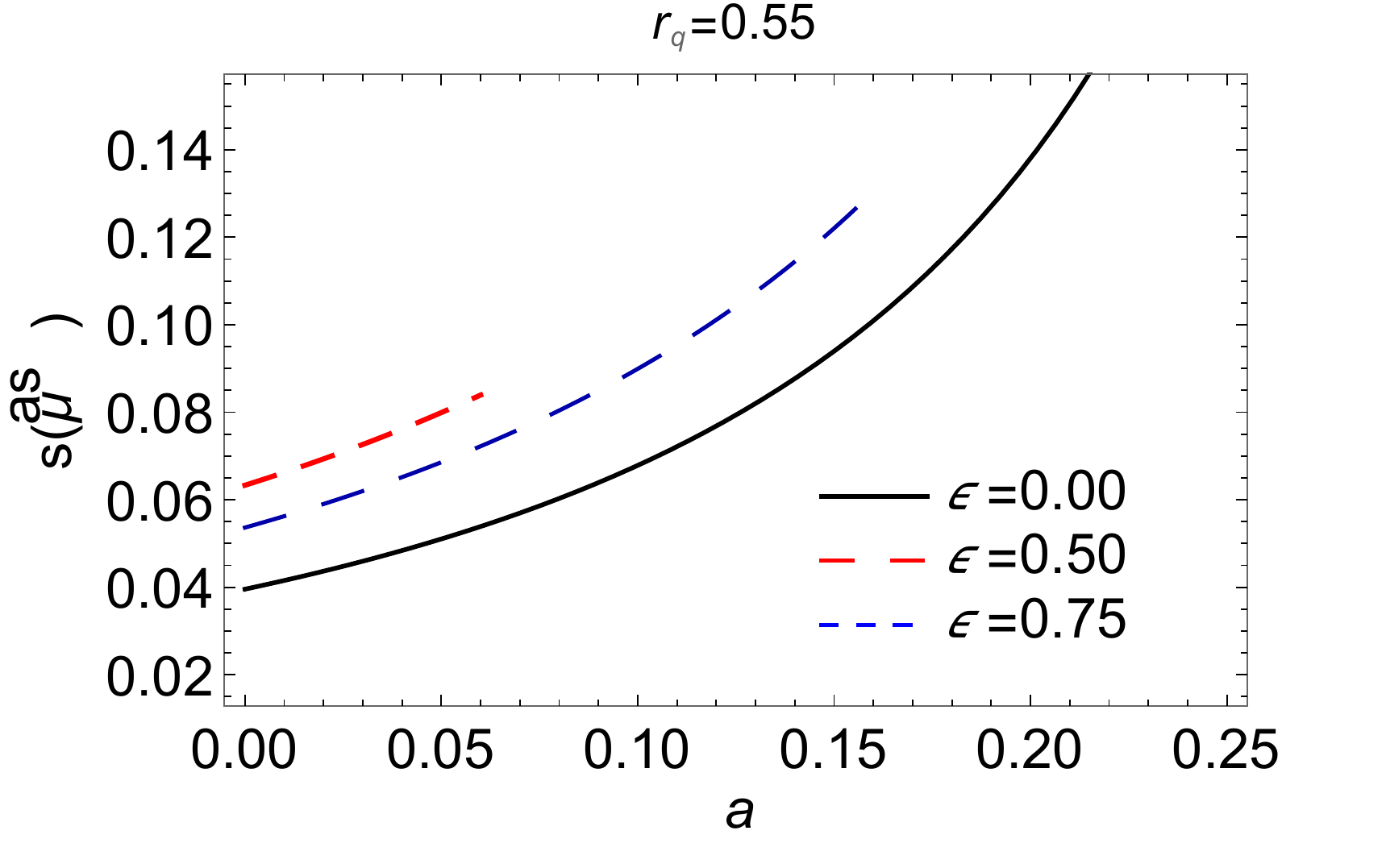}\\
			
			\includegraphics[scale=0.50]{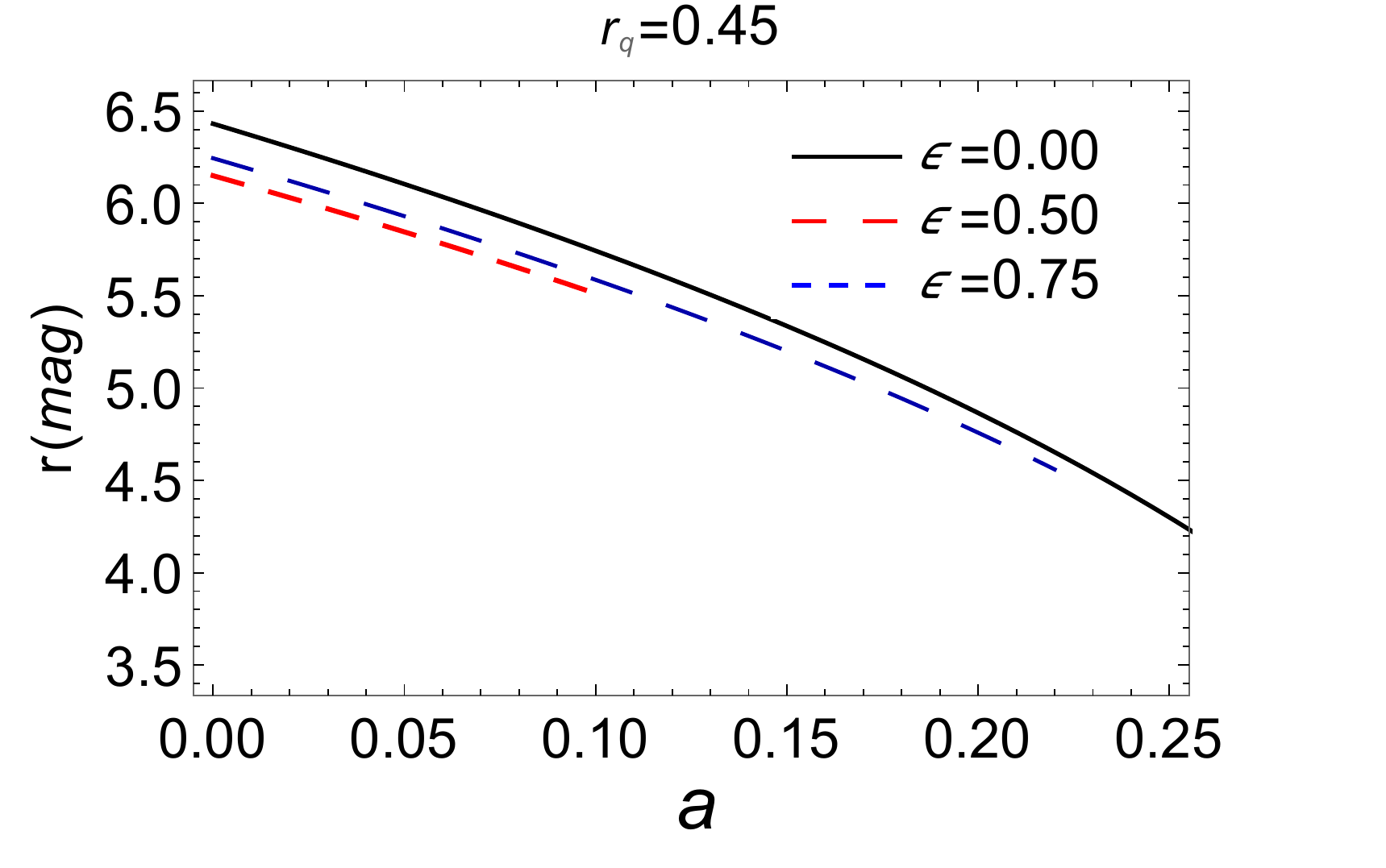}\hspace*{-0.7cm}
		    \includegraphics[scale=0.50]{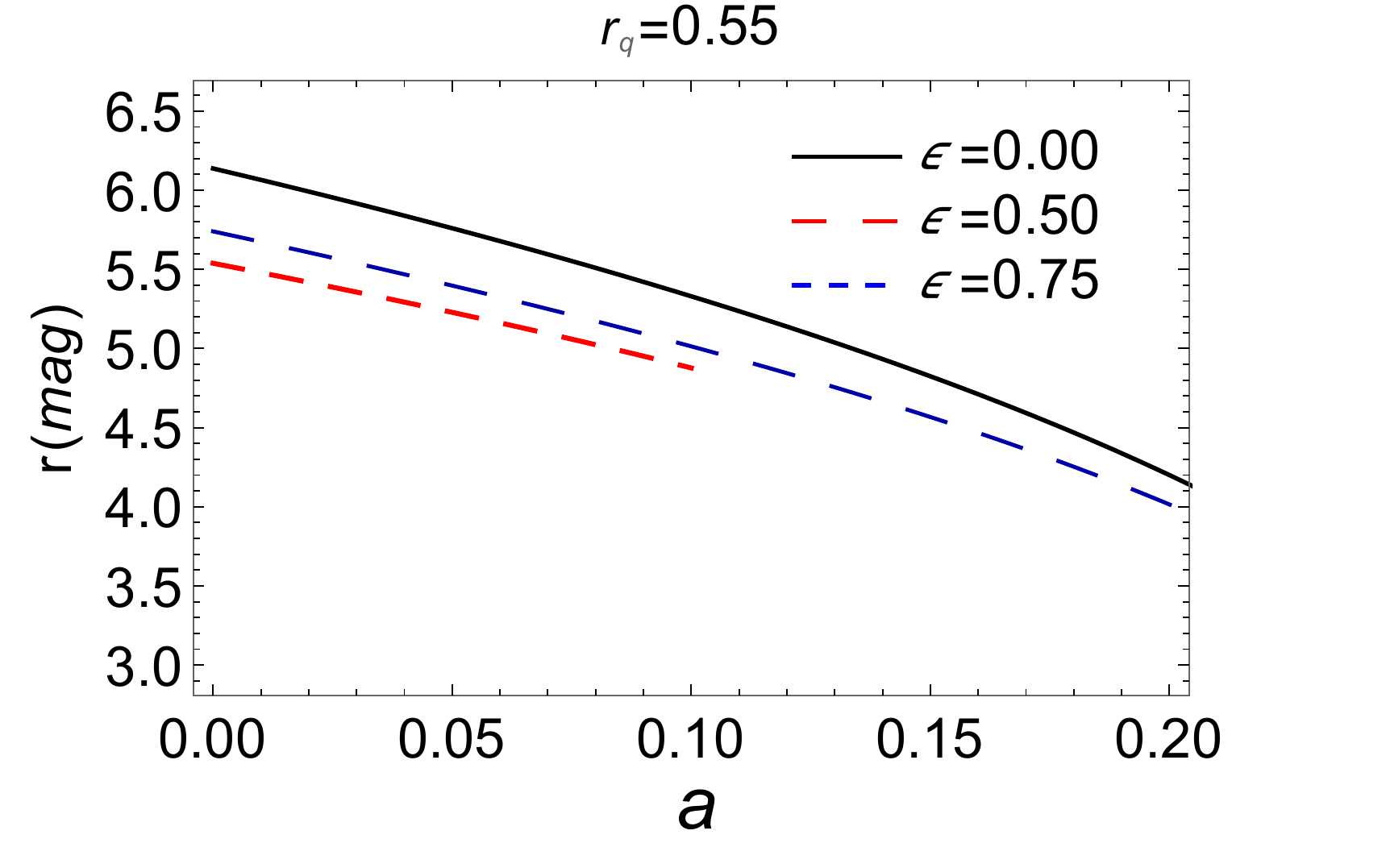}
			\end{tabular}
	\end{centering}
	\caption{Plot showing the variation of lensing observables  $\theta_{\infty}$, $s$, and $r_{mag}$ as a function of $a$ for different values of $r_q$ and $\epsilon$ for M87.}\label{plot1b}		
\end{figure*}  
It is worthy to mention that $\theta_1^E$ is the angular position of the outermost ring. The Fig.~\ref{plotER}, depicts the the angular position $\theta_1^E$ various black holes. Like the Einstein's one of the most important quantities is the image magnification which is viewed as the ratio of solid angles made by the image and the source with the central object, such that for the $n$th image the magnification is defined as \cite{Bozza:2002zj,Bozza:2002af}
\begin{eqnarray}
\mu_n &=& \frac{1}{\beta} \Bigg[\frac{u_m}{D_{OL}}(1+e_n) \Bigg(\frac{D_{OS}}{D_{LS}}\frac{u_me_n}{D_{OL}p}  \Bigg)\Bigg].
\end{eqnarray}
As expected the magnification decreases and the image becomes fainter as $n$ increases. We have from  \cite{Bozza:2002zj}, the important observables describing the rotating black holes in EiBI gravity theory as
\begin{eqnarray}
\theta_\infty &=& \frac{u_m}{D_{OL}}\\
s &=& \theta_1-\theta_\infty \approx \theta_\infty (e^{\frac{q-2\pi}{p}})\\
r_{\text{mag}} &=& \frac{\mu_1}{\sum{_{n=2}^\infty}\mu_n } \approx 
e^{\frac{2 \pi}{p}},
\end{eqnarray} 
where $s$ is the angular separation, $r_{\text{mag}}$ is the ratio of the the flux magnification of the first image and sum of the flux magnifications from other images. We plot these observables in a realistic scenario of various black holes, such as, the Sgr A*, M87 \cite{Kormendy:2013}. We consider $M=4.3\times 10^6 M_{\odot}$ and $d=8.35$~Kpc \cite{Do:2019vob} for Sgr A*, and $M=6.5 \times 10^9 M_{\odot} $ and $d=16.8$~Mpc  for M87\cite{Akiyama:2019eap}.
From the plots in Figs.~6 and 7 it is clear that the angular separation increases but the angular position ($\theta_\infty$) and the flux magnitude ($r_{\rm mag}$) decrease with different values of $r_q$ and $\epsilon$.\\

\begin{table}
\label{tab1}
       \caption{The table shows the set values of the spin parameter $a$ and the full deflection angle, i.e., $(a,2\theta_\infty)$ of rotating EiBI black hole with the charge parameter $r_q/r_S= 0.15$, and  $\epsilon/r_S^2=0.0$ (first column), $\epsilon/r_S^2=0.1$ (second column) and  $\epsilon/r_S^2=0.15$ (third column) as a model to the M87 black holes.}\label{table2}
        \begin{tabular}{|c c c| c c c | c c c|}
        \hline
$a,2\theta_\infty(\mu as)$&  &&$a,2\theta_\infty(\mu as)$& &&$a,2\theta_\infty(\mu as)$&& \\
            \hline          
0.0,42.285&  & &0.0, 42.2752&  &&0.0, 42.2703&&\\
0.0104,42&  & &0.0102,42&  &&0.0084, 42&&\\
0.05,40.6065&  & &0.05, 40.5963&  &&0.05, 40.5912&&\\
0.0963,39&  & &0.0963, 39&  &&0.0960, 39&&\\
0.1,38.8696&  & &0.1, 38.859&  &&0.1, 38.8536&&\\
0.15,37.06241&  & &0.15 , 37.0513&  &&0.15, 37.0457&&\\
0.2,35.1682&  & &0.2,35.1566&  &&0.2, 35.1507&&\\	
0.25,33.1626&  & &0.25,33.1504&  &&0.25, 33.1442&&\\
0.3,31.0074&  & &0.3, 30.9945&  &&0.3, 30.988&&\\
0.35,28.6361&  &&0.35, 28.6225&  &&0.35, 28.6157&&\\
0.4,25.9127&  &&0.4,25.8985&  &&0.4, 25.8914&&\\
0.45,22.4438&  & &0.45, 22.4298&  &&0.45, 22.4228&&\\
            \hline
        \end{tabular}
\end{table}
We tabulated the values of the pair $(a,2\theta_\infty)$ for a fixed value of the rotation parameter $r_q/r_S=0.15$ (one should remember that $r_S=2M$ is the Schwarzschild radius, where $M$ is the mass of the EiBI black hole). We can estimate the value of the deflection angle for M87 black holes using the restricted values of the parameters $r_q$ and $\epsilon$. If we assume the parametric values of these parameters as tabulated in Table~I, we can see that we have the diameter of the photon ring of the M87 black holes as reported by the Event horizon Telescope (EHT) Collaboration. Although, we do not get the upper bound of the diameter of the photon ring,  i.e., $45\,\mu$as, as reported by EHT, but we get the values $42\,\mu$ as and $39\,\mu$ as, as the bound lies as $42\pm 3\,\mu$as (see the tabulated values for the M87 black holes for the reference).   
\begin{table}\label{tab2}
       \caption{The table shows the set values of the spin parameter $a$ and the full deflection angle, i.e., $(a,2\theta_\infty)$ of rotating EiBI black hole with the charge parameter $r_q/r_S= 0.15$, and  $\epsilon/r_S^2=0.0$ (first column), $\epsilon/r_S^2=0.1$ (second column) and  $\epsilon/r_S^2=0.15$ (third column) as a model to the SgrA$^*$.}\label{table2}
        \begin{tabular}{|c c c| c c c | c c c|}
        \hline
$a,2\theta_\infty(\mu as)$&  &&$a,2\theta_\infty(\mu as)$& &&$a,2\theta_\infty(\mu as)$&& \\
            \hline          
0.0,50.6784&  & &0.0, 50.6666&  &&0.0, 50.6608&&\\
 0.05,48.6666&  & &0.05, 48.6544&  &&0.05, 48.6483&&\\
 0.1,46.585&  & &0.1, 46.5722&  &&0.1, 46.5659&&\\
0.1361,45& &&0.136, 45& &&0.1363, 45&&\\
 0.15,44.4191&  & &0.15, 44.4058&  &&0.15, 44.3991&&\\
0.2,42.1489&  & &0.2, 42.1349&  &&0.2, 42.1279&&\\	
 0.2031,42& &&0.202, 42& &&0.204,42&&\\
0.25,39.7452&  & &0.25,39.73057&  &&0.25, 39.723&&\\
 0.2648,39&  & &0.26420, 39&  &&0.265, 39&&\\
0.3,37.1622&  & &0.3, 37.1467&  &&0.3, 37.1389&&\\
0.35,34.3202&  &&0.35,34.3039&  &&0.35,34.2958&&\\
0.4,31.0562&  &&0.4, 31.0392&  &&0.4, 31.0306&&\\
0.45,26.8988&  & &0.45, 26.882&  &&0.45, 26.8736&&\\
 \hline
        \end{tabular}
\end{table}

As a consistency check we have also calculated the diameter of the photon ring of the SgrA$^*$. We see that if we allow the window of $2\theta_\infty=42\pm 3\,\mu$as for the diameter of the photon ring for the SgrA$^*$, then for the particular values of the parameters $r_q$ and $a$, we always have the satisfactory results. We have tabulated them in Table~II.
\section{Weak gravitational lensing}\label{weak_lensing}
In the present section we deal with the weak lensing of the rotating charged black holes in EiBI theory. We rewrite the metric in usual in ($t,r,\theta,\phi$) coordinates so as to get the form
\begin{align}
ds^2=-X(r,\theta)dt^2-2U(r,\theta)dtd\phi+Y(r,\theta)dr^2+Z(r,\theta)d\theta^2+V(r,\theta)d\phi^2 . 
\label{ds2-axial}
\end{align}
where form of $X(r,\theta)$, $Y(r,\theta)$, $Z(r,\theta)$, $V(r,\theta)$ and $U(r,\theta)$ can be seen when we compare the metric (\ref{eq:axialline}) with  (\ref{ds2-axial}). We consider only the null rays for the propagation, which is seen by computing $ds^2 = 0$ for $dt$ as
\begin{align}
dt=& \sqrt{\gamma_{ij} dx^i dx^j} +N_i dx^i , 
\label{opt} 
\end{align}
where 
$i, j=1,2, 3$, and 
$\gamma_{ij}$ and $N_i$ are defined accordingly as 
\begin{align}
\gamma_{ij}dx^idx^j \equiv&
\frac{Y(r,\theta)}{X(r,\theta)}dr^2
+\frac{Z(r,\theta)}{X(r,\theta)}d\theta^2
+\frac{X(r,\theta)V(r,\theta)+U^2(r,\theta)}{A^2(r,\theta)}d\phi^2 , 
\label{gamma}
\\
N_idx^i \equiv& -\frac{U(r,\theta)}{X(r,\theta)} d\phi . 
\label{beta}
\end{align}
The properties of the $\gamma^{ij}$ is followed from the relation 
$\gamma^{ij}\gamma_{jk} = \delta^i_{~k}$. $\gamma_{ij}$ encodes the properties of a three-dimensional Riemannian space in which the trajectories of the null rays are described by the motion along a spatial curve.

Now we use the metric (\ref{gamma}) and then the Gauss-Bonnet theorem to have the definition of the light deflection angle which described as \cite{Gibbons:2008rj, Ishihara:2016vdc, Ishihara:2016sfv} 
\begin{eqnarray}
\alpha_D=-\int\int_{{}_O^{\infty}\Box_{S}^{\infty}} K dS+\int_{S}^{O} k_g dl,\label{deflectionangle}
\end{eqnarray}
where $K$ is the curvature of the three-surface along which light propagates, $k_g$ is the geodesics curvature of the light curves, $dS$ is the area element, and $dl$ is the line element. We define the curvature of the three surface at the equatorial plane ($\theta=\pi/2$), as follows
\begin{eqnarray}
\label{gaussian_curve1}
K&=&\frac{{}^{3}R_{r\phi r\phi}}{\gamma},\nonumber\\
&=&\frac{1}{\sqrt{\gamma}}\left(\frac{\partial}{\partial \phi}\left(\frac{\sqrt{\gamma}}{\gamma_{rr}}{}^{(3)}\Gamma^{\phi}_{rr}\right) - \frac{\partial}{\partial r}\left(\frac{\sqrt{\gamma}}{\gamma_{rr}}{}^{(3)}\Gamma^{\phi}_{r\phi}\right)\right),
\end{eqnarray}
where $\gamma$ is the determinant of the metric when $\theta=\pi/2$. For the rotating axially symmertic spacetime, equation (\ref{gaussian_curve1}) becomes \cite{Ishihara:2016vdc, Ishihara:2016sfv}
\begin{eqnarray}
\label{gaussian_curve2}
K&=&-\sqrt{\frac{X^3}{Y(XV+U^2)}}\frac{\partial}{\partial{r}}\left[\sqrt{\frac{X^3}{Y(XV+U^2)}}\frac{\partial}{\partial{r}}\left(\frac{XV+U^2}{X^2}\right)\right]
\end{eqnarray}
Therefore $K$ is evaluated to be
\begin{eqnarray}
K&=&\left(\frac{3r_q^2}{2x^4}+\frac{3\epsilon r_q^2+r_q^2 a^2}{x^6} \right)-\left(\frac{1}{x^3}+\frac{6a^2+3r_q^2+2\epsilon r_q^2}{x^5} \right)r_S+\left(\frac{3}{4x^4}+\frac{-6a^2+\epsilon r_q^2+5r_q^2}{2x^6}\right)r_S^2\nonumber\\
&&+\mathcal{O}\left( \frac{a^2\epsilon r_S^2}{x^8},\frac{r_q^2 \epsilon^2r_S^2}{x^8},\frac{a^2 r_q^2 r_S^2}{x^8},\frac{r_S^3}{x^5}\right).
\end{eqnarray}
It is clear that to calculate the leading order contribution we approximated the calculations to the weak field limit and all the higher order terms are safely ignored. The Gaussian curvature is integrated over the quadrilateral which is closed so that \cite{Ono:2017pie}
\begin{equation}
\int\int_{{}_O^{\infty}\Box_{S}^{\infty}} K dS= \int_{\phi_S}^{\phi_O}\int_{\infty}^{x_0} K \sqrt{\gamma}dr d\phi,\label{Gaussian}
\end{equation}
where $x_0$ is the closest distance to the black hole. On  utilizing Eqs.~(\ref{phidot}) and (\ref{xdot}) and choosing $u=1/x$, we can express the equation for light orbit as follows 
\begin{equation}
\left(\frac{du}{d\phi}\right)^2=F(u),\label{orbit}
\end{equation}
with 
\begin{equation} F(u)=\frac{u^4
\left(XU+V^2\right)\left(U-2Vb-Xb^2\right)}{\Big(Y(V+Xb)\Big)^2},
\end{equation}
with $b\equiv\ell/\mathcal{E}$ is defined as the impact parameter. Using the weak field solution $u=(\sin\phi)/b +\mathcal{O}(r_S,r_S^2)$ \cite{Ono:2017pie}, Eq.~(\ref{Gaussian}) reduces to
\begin{equation}
\label{Kds}
\int\int_{{}_O^{\infty}\Box_{S}^{\infty}} K dS= \int_{\phi_S}^{\phi_O}\int_{0}^{\frac{\sin\phi}{b}}-\frac{K\sqrt{\gamma}}{u^2}du d\phi.
\end{equation}
Therefore for the metric (\ref{gamma}), the integral (\ref{Kds}) reads as
\begin{eqnarray}
\int\int K dS&=&\frac{r_S}{b}\left(\sqrt{1-b^2u_O^2}+\sqrt{1-b^2u_S^2}\right) +\frac{r_Sa^2}{3b^3}\left((2+b^2u_S^2)\sqrt{1-b^2u_S^2}+(2+b^2u_O^2)\sqrt{1-b^2u_O^2} \right)\nonumber\\
&-& \frac{r_Sr_q^2}{3b^3}\left((16+b^2u_S^2)\sqrt{1-b^2u_S^2} +(16+b^2u_O^2)\sqrt{1-b^2u_O^2}  \right)\nonumber\\
&-&\frac{11a^2r_Sr_q^2}{25b^5}\left((3b^4u_S^4+4b^2u_S^2+8)\sqrt{1-b^2u_S^2}+(3b^4u_O^4+4b^2u_O^2+8)\sqrt{1-b^2u_O^2} \right)\nonumber\\
&-&\left(\frac{3r_q^2}{8b} +\frac{3r_S^2}{16b}\right)\left(u_S\sqrt{1-b^2u_S^2}+u_O\sqrt{1-b^2u_O^2}\right)\nonumber\\
&-&\left(\cos^{-1}bu_S+\cos^{-1}bu_O\right)\left(\frac{3r_q^2}{8b^2} +\frac{3a^2r_q^2}{8b^4} -\frac{15r_S^2}{16b^2}-\frac{9r_S^2a^2}{16b^4}-\frac{9\epsilon r_q^2}{16b^4}-\frac{27r_S^2r_q^2}{256b^4}\right)\nonumber\\
&+&\left(-\frac{a^2r_q^2}{8b^3}-\frac{15r_S^2a^2}{16b^3}-\frac{3\epsilon r_q^2}{8b^3}\right)\left(u_S(3+2b^2u_S^2)\sqrt{1-b^2u_S^2}+u_O(3+2b^2u_O^2)\sqrt{1-b^2u_O^2}\right)\nonumber\\
&+&\mathcal{O}\left( \frac{a^2r_q^2r_S^2}{b^6},\frac{r_q^2r_S^2\epsilon}{b^6},\frac{r_S^3}{b^3}\right),\label{KdS1}
\end{eqnarray}
where $u_O$ and $u_S$ are defined, respectively, as $\cos\phi_o=-\sqrt{1-b^2u_O^2}, \cos\phi_s=\sqrt{1-b^2u_S^2}$. The geodesic curvature of the manifold $^{(3)}\mathcal{M}$ is defined to be \cite{Ono:2017pie}
\begin{equation}
k_g=-\frac{1}{\sqrt{\gamma\gamma^{\theta\theta}}}N_{\phi,r},
\end{equation}
which reflects the fact the non-rotating black hole does not contribute to it, thereby making a crucial contribution to the light deflection angle. Hence, the geodesic curvature for metric (\ref{gamma}) reads
\begin{equation}
k_g=\left({\frac {a}{2{x}^{4}}}-{\frac {3a{r_{{q}}}^{2}}{4{x}^{6}}} \right) {r_{{S}}}^{2}+ \left( {\frac {a}{{x}^{3}}}-{\frac {3a{r
_{{q}}}^{2}}{4{x}^{5}}} \right) r_{{S}}-{\frac {a{r_{{q}}}^{2}}{{x}^{4}
}}-{\frac {4a\epsilon\,{r_{{q}}}^{2}}{{x}^{6}}}
+\mathcal{O}\left(\frac{r_S^3a}{x^5},\frac{ar_S^2r_q^2\epsilon}{x^8}\right).
\end{equation}
We consider a coordinate system which is centered at the position of the lens, we can take the approximation of the the light curve such that $r=b/{\cos\vartheta}$ and $l=b\tan\vartheta$ \cite{Ono:2017pie}. Therefore, the geodesic curvature in its path integral form reads 
\begin{eqnarray}
\int_S^O k_g dl &=& \int_S^O\Bigg(\left(\frac{a}{b^2}\cos\theta-\frac{3}{4}\frac{ar_q^2}{b^4}\cos^3\theta\right)r_S+\left(\frac{a}{2b^3}\cos^2\theta-\frac{3}{4}\frac{ar_q^2}{b^5}\cos^4\theta\right)r_S^2-\frac{ar_q^2}{b^3}\cos^2\theta\nonumber\\
&-&\frac{4a\epsilon r_q^2}{b^5}\cos^4\theta\Bigg)d\theta+\mathcal{O}\left(\frac{r_S^3a}{b^4}\right)\nonumber\\
&=&-\frac{r_Sa}{b^2}\left(\sqrt{1-b^2u_S^2}+\sqrt{1-b^2u_O^2}\right)\nonumber\\
&&+\left(\frac{ar_q^2}{2b^2}-\frac{r_S^2a}{b^2}+\frac{9ar_qr_S^2}{32b^4}+\frac{3}{2}\frac{\epsilon ar_q^2}{b^4}\right)\left(u_S\sqrt{1-b^2u_S^2}+u_O\sqrt{1-b^2u_O^2}\right)\nonumber\\
&&+\left(\frac{ar_q^2}{2b^3}-\frac{r_S^2a}{4b^3}-\frac{45ar_q^2r_S^2}{32b^5}+\frac{3}{2}\frac{a\epsilon r_q^2}{b^5}\right)\left(\cos^{-1}bu_O+\cos^{-1}bu_S\right)\nonumber\\
&&-\frac{15ar_q^2r_S^2}{32b^4}\left(u_S(3+2b^2u_S^2)\sqrt{1-b^2u_S^2}+u_O(3+2b^2u_O^2)\sqrt{1-b^2u_O^2}\right)\nonumber\\
&&+ \frac{\epsilon a r_q^2}{2b^4}\left(u_S(3+2b^2u_S^2)\sqrt{1-b^2u_S^2}+u_O(3+2b^2u_O^2)\sqrt{1-b^2u_O^2}\right) +\mathcal{O}\left(\frac{r_S^3a}{b^4},\frac{r_S^4ar_q^2}{b^7}\right),
\label{geodesiccurvature}
\end{eqnarray}   
In deriving the above expression we adopted the prograde motion along the null geodesics ($dl>0$), while for $dl<0$, we obtain the retrograde motion with the extra terms with `-' sign. For asymptotically large distances we have, $u_O\to 0$ and $u_S\to 0$, so the deflection angle gives us the expression
\begin{eqnarray}
\alpha_D&=&\left(\frac{a\pi r_q^2}{2b^3}-\frac{3\pi r_q^2}{8b^2}-\frac{3\pi a^2 r_q^2}{8b^4}-\frac{9\epsilon\pi r_q^2}{16b^4}-\frac{3\epsilon a\pi r_q^2}{2b^5}\right) + \left(\frac{4}{b} -\frac{4a}{b^2}+\frac{8a^2}{3b^3}+ \frac{32r_q^2}{6b^3}+\frac{176a^2r_q^2}{50b^5}\right)r_S\nonumber\\
&&+\left( \frac{15\pi}{16b^2}-\frac{a\pi}{b^3}+\frac{9\pi a^2}{16b^4}+\frac{27\pi r_q^2}{256 b^4}-\frac{45\pi ar_q^2}{32b^5}\right)r_S^2 +\mathcal{O}\left(\frac{r_S^3}{b^3}, \frac{r_S^4}{b^4}\right),~\label{deflection}
\end{eqnarray}
which reduces the Kerr-Newman limit of the deflection angle \cite{Kumar:2019ku} when $\epsilon\to 0$, such that 
\begin{eqnarray}
\alpha_D^{\text{KN}}&=&\left(\frac{a\pi r_q^2}{2b^3}-\frac{3\pi r_q^2}{8b^2}-\frac{3\pi a^2 r_q^2}{8b^4}\right) + \left(\frac{4}{b} -\frac{4a}{b^2}+\frac{8a^2}{3b^3}+ \frac{32r_q^2}{6b^3}+\frac{176a^2r_q^2}{50b^5}\right)r_S\nonumber\\
&&+\left( \frac{15\pi}{16b^2}-\frac{a\pi}{b^3}+\frac{9\pi a^2}{16b^4}+\frac{27\pi r_q^2}{256 b^4}-\frac{45\pi ar_q^2}{32b^5}\right)r_S^2 +\mathcal{O}\left(\frac{r_S^3}{b^3}, \frac{r_S^4}{b^4}\right),
\end{eqnarray}
which in addition  for $r_q\to 0$, reduces to the expression for deflection angle for Kerr black holes \cite{Ishihara:2016vdc, Ishihara:2016sfv} 
\begin{equation}
\alpha_D^{\text{Kerr}}=\left(\frac{4}{b} -\frac{4a}{b^2}+\frac{8a^2}{3b^3}\right)M+\left( \frac{15\pi}{4b^2}-\frac{a\pi}{b^3}+\frac{9\pi a^2}{4b^4}\right)M^2 +\mathcal{O}\left(\frac{M^3}{b^3}, \frac{M^4}{b^4}\right),
\end{equation}
For the nonrotating ($a=0$) black hole in EiBI theory, the deflection angle has the form
\begin{eqnarray}
\alpha_D&=&\left(-\frac{3\pi r_q^2}{8b^2}-\frac{9\epsilon\pi r_q^2}{16b^4}\right) + \left(\frac{4}{b}+ \frac{32r_q^2}{6b^3}\right)r_S+\left( \frac{15\pi}{16b^2}+\frac{27\pi r_q^2}{256 b^4}\right)r_S^2 +\mathcal{O}\left(\frac{r_S^3}{b^3}, \frac{r_S^4}{b^4}\right),~\label{deflection}
\end{eqnarray}

\section{Conclusions}\label{Sec6}
The general theory of relativity has been tested and the theory incredibly matches with the local astrophysical evidences. The black holes are one of the strangest objects that were predicted in general relativity but still there are only few concepts which have been verified on the experimental level. The experimental discovery that the black hole solutions such as Schwarzschild and Kerr metrics, are not the actual real black holes would have pointed that a strong-field deviation from general relativity may have deep implications at the fundamental level. In the present paper, we investigated the gravitational lensing in the strong field approximation of the black holes in EiBI theory. Using the standard procedure for calculating the impact parameter, we study numerically the total azimuthal deflection $\alpha_D$ of light rays. We find that the charge parameter $r_q$ and the Born-Infeld parameter $\epsilon$ influences the null geodesics. The coefficients $p$ and $q$ also have been obtained and plotted numerically which show that with fixed values of the charge $r_q$, the coefficient $p$ increases with rotation parameter $a$, while $q$ is decreasing with $a$. Figure \ref{plot2} shows that the coefficients $p$ and $q$ share the same property as those of Kerr-Newman ($\epsilon=0$) and the stationary axially symmetric black holes in EiBI theory ($\epsilon\neq 0$). The deflection angle $\alpha_D$ showed the monotonic behaviour with the rotation parameter $a$ and it diverges at $u=u_m$ which have been shown with dots on the horizon lines in Figure \ref{plot4}. As an application to the realistic scenario, we calculated the strong lensing observables $s$, $\theta_\infty$, and $r_{\text{mg}}$ for  SgrA* and M87 black holes. 

We further calculate the weak field gravitational lensing for the rotating black holes in EiBI theory. We have shown that in the limit $\epsilon\to\infty$, our results macth with the Kerr-Newman black holes. This way all limit cases of Kerr black holes ($r_q=0$) are satisfied. Our results may be important from phenomenological point of view as the results from EHT slightly deviates from the Kerr black holes. This way we can implement our investigations to the study of the astrophysical scenario. 

\section{Acknowledgments} 
MSA's research is supported by the ISIRD grant 9-252/2016/IITRPR/708. The authors would like to thank Mr.~Arpit Maurya for useful discussions and critical comments on the revised version of the manuscript.

\end{document}